\documentclass[useAMS,usenatbib,usegraphicx]{mn2e}
\usepackage{longtable}
\def\etal{{et al.}}
\def\etals{et al.}

\def\la{\mathrel{\hbox{\rlap{\hbox{\lower4pt\hbox{$\sim$}}}\hbox{$<$}}}}
\def\gt{\mathrel{\hbox{$>$}}}

\def\ga{\mathrel{\hbox{\rlap{\hbox{\lower4pt\hbox{$\sim$}}}\hbox{$>$}}}}
\def\lt{\mathrel{\hbox{$<$}}}

\title[The near infrared luminosity function of the Norma cluster]{The Norma cluster (ACO3627): II. The near infrared $K_s$-band luminosity function}
\author[R. E. Skelton, P. A. Woudt and R. C. Kraan-Korteweg]{R. E. Skelton$^{1,2}$\thanks{Email: skelton@mpia-hd.mpg.de}, P. A. Woudt$^{1}$ and R. C. Kraan-Korteweg$^{1}$\\
$^{1}$Department of Astronomy, University of Cape Town, Private Bag X3, Rondebosch, 7701, South Africa\\
$^{2}$Max-Planck Institute for Astronomy, K\"onigstuhl 17, Heidelberg, 69117, Germany }

\begin{document}

\date{Accepted 2009 April 14. Received 2009 April 07; in original form 2008 January 29}

\pagerange{\pageref{firstpage}--\pageref{lastpage}} \pubyear{2009}

\maketitle

\label{firstpage}

\begin{abstract}
A deep $K_s$-band photometric catalogue of galaxies at the core of the
rich, nearby Norma cluster (ACO3627) is presented. The survey covers
about $45$\arcmin$\times 45$\arcmin (slightly less than 
1/3 Abell radius), which corresponds to $\sim$
0.8$h_{70}^{-2}$~Mpc$^2$ at the adopted distance ($v_{CMB}/H_0$) of
70~$h_{70}^{-1}$\,Mpc of this cluster. The survey is estimated to be
complete to a magnitude of $M_{K_s} \la -19.5$~mag. This extends into
the dwarf regime, 6 magnitudes below $M_{K_s}^*$. The catalogue
contains 390 objects, 235 of which are classified as likely or
definite galaxies and 155 as candidate galaxies.
The $K_s$-band luminosity function (LF) is
constructed from the photometric sample, using a spectroscopic
subsample to correct for fore- and background
contamination.  We fit a Schechter function with a characteristic magnitude
of $M^*_{K_s}$ = -25.39 $\pm$ 0.80 mag and faint-end slope of $\alpha$ = -1.26 $\pm$ 0.10 
to the data. The shape of
the LF is similar to those found in previous determinations of the
cluster LF, in both optical and near infrared. The Schechter parameters 
agree well with those of recent field LFs, suggesting that both the
shape of the bright end and the faint end slope are relatively
insensitive to environment.

\end{abstract}


\section{INTRODUCTION}

The Norma cluster (ACO 3627; \citealt{Abell89}) at ($l$, $b$, $v$) =
(325$^{\circ}$, -7$^{\circ}$, 4844 km s$^{-1}$) is the nearest rich,
massive cluster of galaxies (\citealt{Kraan-Korteweg96};
\citealt{Woudt98}), with properties comparable to the Coma cluster
\citep{Mazure98}. It has remained relatively unexplored in comparison
to its well known counterpart, mainly because of its location at low
Galactic latitudes in the southern Zone of Avoidance (ZOA) where dust
extinction and star crowding dilute its appearance. The Norma cluster's
prominence -- and importance with regard to the dynamics
of the local Universe -- was only realized in the late 1990s (\citealt
{Kraan-Korteweg96}; \citealt{Boehringer96}).

This paper presents a deep near-infrared $K_s$-band survey of the central part
of the cluster. It is the second in a series of papers on the properties of the
Norma cluster. Paper I was dedicated to a dynamical analysis of the cluster 
\citep{Woudt08}. An introduction to the cluster and its relevance to the 
Great Attractor overdensity and local cosmic flow fields is discussed there.

Observations in the near infrared (NIR) wavelength range are useful
for penetrating the ZOA: dust extinction is up to an order of
magnitude smaller in the NIR than in the optical. The NIR has further
advantages. While star-forming regions are bright at optical wavelengths,
making optical observations particularly sensitive to galaxies where
star formation is taking place, the NIR is much less
affected by short bursts of star formation and therefore a better
indicator of galaxy mass \citep{Andreon00a}. The NIR has the advantage
of high resolution and sensitivity to both spiral and elliptical
galaxies, compared to longer wavelength regimes where instruments have
limited resolution and are sensitive to nearby late-type galaxies. As
a result, the NIR provides information on the underlying mass
distribution of galaxies. An additional consideration for evolutionary
studies is that $k$-corrections in the NIR are smaller than in the
optical and less dependent on galaxy type. The NIR is thus useful for
comparisons over a range of redshifts.

The deep $K_s$-band images obtained for this investigation cover the
central 0.8~Mpc$^2$ of the Norma cluster. The survey is
close to 3 magnitudes deeper than the 2 Micron All Sky Survey (2MASS,
\citealt{Skrutskie06}) and has much better seeing and spatial resolution (0.45
arcsec pixels compared to 2 arcsec pixels). The latter is particularly
important since the combination of low resolution and high star
density creates a ``NIR'' Zone of Avoidance which affects the
completeness of the 2MASS Extended Source Catalogue (XSC) at the location of the Norma cluster
(\citealt{Kra05}; \citealt{Kra05a}). 

Based on the survey, the $K_s$-band luminosity function (LF hereafter)
for the cluster is determined. The LF is a probability distribution,
describing the relative numbers of galaxies of different magnitudes
within a particular environment. It is a fundamental tool for
observational cosmology and the study of galaxy formation and
evolution \citep{Binggeli88}. Both the local field LF and those of
nearby clusters are used for comparison with higher redshift samples
to study the evolution of galaxies. 

The Coma cluster has been used as the archetypal rich zero-redshift cluster in a number of LF studies, dating back to the 1980s (see for e.g. \citealt{Thompson80}; \citealt{Andreon00b}). The Norma cluster now provides an ideal opportunity to test and complement existing knowledge on local clusters, due to the recent development of techniques to deal with ZOA data (e.g. automated star-subtraction with routines such as \textsc{killall}; \citealt{Buta99}) and infrared array detectors that can image large areas in the NIR. The NIR $K_s$-band LF determined for the Norma cluster can be used as a comparitive tool to examine changes in cluster composition -- both locally and with increasing redshift -- and thus explore galaxy evolution in dense environments. 


An introduction to LFs and summary of recent determinations for clusters and the field are given in Section~2. Section~3 provides details of the NIR observations of the Norma cluster and describes the data reduction process, including star-subtraction and photometry. In Section~4, the results of the survey are presented. The $K_s$-band catalogue is described and cross-referenced with the optical catalogue of \citet{Woudt01} and the distribution of the objects and their properties are examined. The LF is derived in Section 5 and compared to those of other clusters and environments. A Hubble constant of $H_0 = 70$~km~s$^{-1}$Mpc$^{-1}$ is used throughout this work.

\section{LUMINOSITY FUNCTIONS} \label{LFsec}

The general LF $\Phi(M)$ is a probability distribution over absolute
magnitude ($M$) \citep{Binggeli88}. It can usually be well described 
by the function 
\begin{equation}\label{schech_fnM}
\phi(M) dM = \phi^* 10^{0.4(\alpha + 1)(M^* - M)}e^{{-10^{0.4(M^* - M)}}} dM
\end{equation}
proposed by \citet{Schechter76}. $M^*$, $\alpha$ and $\phi^*$ are free
parameters representing the characteristic magnitude (the ``knee'' in
the curve), the faint end slope and the normalisation,
respectively. The Schechter function is essentially a power-law at the
faint end and an exponential at the bright end. The Schechter function
parameters are found to be remarkably similar for the field and
cluster environments, despite the differences in density and relative
fractions of morphological types. 

Not all clusters can be modelled by a Schechter function, however. The
Virgo cluster, for example, has a LF with a ``double wave'' structure
\citep{Binggeli88}. Clusters containing cD galaxies seem to have a
steeper bright end than those without cD galaxies. Often the brightest
galaxy is excluded to obtain a better fit, particularly in cD
clusters. The Schechter parameters have also been found to vary depending on the
region sampled within the cluster, the size of the area and the
limiting magnitude \citep{Andreon01}. 

The relatively recent development of wide-field arrays such as the NICMOS (256 $\times$ 256) and HAWAII (1024 $\times$ 1024) HgCdTe detectors enabled LFs in the near infrared to be determined over much larger areas than was previously possible. Up until the mid 90s, deep surveys were only possibly for small areas, making it difficult to determine a local field LF in the NIR \citep{Huang97} or image large enough areas over nearby clusters. The first determinations of the NIR field LF using a wide-field NIR selected survey were made by \citet{Gardner97} and \citet{Szokoly98}. 

The number of galaxies with NIR photometry and redshifts increased by an order of magnitude with 2MASS \citep{Skrutskie06}, making it possible to determine a NIR LF for a sample comparable to the optical, for the first time. \citet{Cole01} and \citet{Kochanek01} determined NIR LFs based on 2MASS Second Instrumental Release data, while \citet{Bell03} used Sloan Digital Sky Survey (SDSS) and 2MASS XSC to find optical and NIR luminosity functions. \citet{Heath Jones06} have improved on this yet further, using 2MASS photometry and a subset of the 6-degree Field Galaxy Survey (6dFGS). The characteristic magnitude obtained was -24.60 $\pm$ 0.03 mag, with a faint end slope of -1.16 $\pm$ 0.04. Their findings suggest that a Schechter function is not an ideal fit to the LF, although the parameters obtained agree well with previous determinations.

\begin{table*}
\begin{minipage}{126mm}
\caption{\label{clusterLF.tbl}Table of Previous NIR Cluster Luminosity Functions}

\begin{tabular}{lllccl}
\hline
Cluster & Reference  & Band &  {$M^*$} &  {$\alpha$} & Sample Notes \\
 & & & {[mag]} & &  \\
\hline
Coma & Mobasher \& Trentham 1998 & $K$ &  & -1.41$^{+0.34}_{ –- 0.37}$ & 0.03 Mpc$^2$, -19.3 $< M_K <$ -16.3\\
Coma & de Propris et al. 1998 & $H$ & -23.82 & -0.93 & central 0.53 Mpc$^2$, $H <$ 14.5\\
Coma & de Propris et al. 1998 & $K$ & -24.02 & -0.98 & central 0.53 Mpc$^2$, $H <$ 14.5\\
Coma & de Propris et al. 1998 & $H$ &  & -1.73 $\pm$ 0.14 & central 0.53 Mpc$^2$, 14 $< H <$ 16\\
Coma & Andreon \& Pello 2000 & $H$  & -23.86 & -1.3 & off-centre 0.30 Mpc$^2$, $H <$ 17\\
AC118 & Barger 1996 &  $K^{'}$ & -24.5 & -1.0 $\pm$ 0.12 & Centre\\
AC118 & Andreon 2001 & $K_s$ & -25.26 & -1.2 & 1.36 Mpc$^2$\\
AC118 & Andreon 2001 & $K_s$ & -23.96 & -0.5 & 0.26 Mpc$^2$ Main clump\\
AC118 & Andreon 2001 & $K_s$ & -23.56 & -0.9 & 0.26 Mpc$^2$ NW clump\\
5 clusters & Trentham \& Mobasher 1998 & $K$ & & -1.38 $\pm$ 0.24 & Composite LF\\	
\hline

\end{tabular}

\footnotesize{\textsc{Note.--} $M^*$ has been scaled for H$_0$ = 70~km~s$^{-1}$~Mpc$^{-1}$. Where no value of $M^*$ is given, the parameter was not constrained in a fit to the Schechter function. No corrections have been made to account for different types of magnitudes used, however characteristic magnitudes can be compared through the colour differences of $H - K \sim 0.22$~mag (see Section~\ref{coma_comp}) and $K^{'}-K\sim0.22(H - K)$~mag \citep{Wainscoat92}.}
\end{minipage}
\end{table*}

Table~\ref{clusterLF.tbl} summarises the most recent cluster LF determinations in the NIR. Here $M^*$ refers to the characteristic magnitude in the wavelength band given in Column 3. The Coma cluster is the most comprehensively studied, in both the optical and NIR bands. The NIR LF has been found to be consistent with the optical LF, shifted by the mean colour in magnitudes (\citealt{Mobasher98}; \citealt{Andreon00a}). A number of authors have found a dip in the optical LF at $B \sim -18$ mag (see for e.g. \citealt{Mobasher03}), corresponding to the transition between giant and dwarf galaxies. \citet{Jerjen97} find the dip to be prominent in the Centaurus and Fornax LFs but less so for Virgo. \citet{Mobasher03} suggest that the dip may be caused by the contribution of the early-type galaxies at the bright end, which could be better fit by a Gaussian, while the faint end could be fit by a power law. This is supported by observations of a similar dip in the NIR LF of Coma at $H \sim -22$~mag \citep{Andreon00a}, which suggests that it is not the result of increased star formation at the bright end due to the hostile cluster environment, but rather due to the morphological mix of galaxies. 

The Schechter parameters depend on the location and size of the area surveyed and the depth of the survey. This can be seen from the variation in the parameters, for example those of \citet{Andreon01} in Table~\ref{clusterLF.tbl}, where different regions of the same cluster have been analysed using the same techniques. \citet{Andreon01} find that the slope of the LF is shallower in higher density regions towards the centre of the cluster, because the relative number of bright galaxies increases while the number of dwarfs decreases. The range in parameters suggests that useful comparisons between cluster LFs can only be made when corresponding areas are surveyed to similar depth(\citet{Jerjen97}).

\section{OBSERVATIONS AND DATA REDUCTION} \label{obs_sec}

A survey of the Norma cluster in the near infrared was undertaken with the Simultaneous 3 colour InfraRed Imager for Unbiased Surveys (SIRIUS) on the Infrared Survey Facility (IRSF), the 1.4-m Japanese telescope at the South African Astronomical Observatory (SAAO) site in Sutherland. SIRIUS allows simultaneous observations in the $J$ ($\lambda =$ 1.26 $\mu m$), $H$ ($\lambda =$ 1.63 $\mu m$) and $K_{s}$ ($\lambda =$ 2.14 $\mu m$) bands. It has three 1024 $\times$ 1024 HgCdTe (HAWAII) arrays. The field of view is approximately 7.8 $\times $ 7.8 arcmin$^{2}$ and the pixel scale is 0.45 arcsec per pixel \citep{Nagashima99,Nagayama03}.  

\begin{figure}
	\begin{center}
		\resizebox{8.4cm}{7.9cm}{\includegraphics{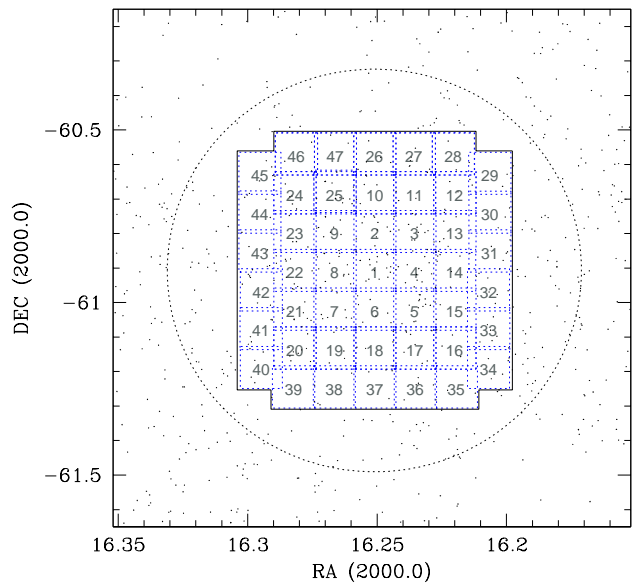}}
	\end{center}
\caption[The 47 IRSF fields centered on the Norma cluster, in equatorial coordinates]{\label{obs_fig}The 47 IRSF fields centered on the Norma cluster, in equatorial coordinates. Each field is 7.8 $\times $ 7.8 arcmin$^{2}$. The surveyed area is demarcated by the solid line. The dotted circle indicates $\frac{1}{3}$ Abell radius (0.6$^\circ$). The dots represent galaxies found in a deep optical search (\citealt{Woudt98}; \citealt{Woudt01}). They identified 106 galaxies within the survey area.}

\end{figure}

Forty seven slightly overlapping fields were observed during three observing runs in June and July 2001 and July 2002, covering an area within 1/3 Abell radius of the cluster (approximately 0.6$^\circ$; \citealt{Woudt05}). Thirty dithered 20-second exposures (10 offset positions along a 30 arcsec circle from the central position) were combined for each field, giving a total exposure time of 600 seconds per frame, in each of the three bands. Near infrared standard stars \citep{Persson98} were observed for photometric calibration. The seeing varied from approximately 1.1 to 1.6 arcsec with a mean full width at half maximum (FWHM) of 1.35 arcsec \citep{Woudt05}. Dark current subtraction, sky subtraction and flat-fielding were done using tasks in the Image Reduction and Analysis Facility\footnote{IRAF is distributed by the National Optical Astronomy Observatory, which is operated by the Association of Universities for Research in Astronomy, Inc., under cooperative agreement with the National Science Foundation} (IRAF) and the data reduction pipeline written by Yasushi Nakashima\footnote{This pipeline is available as the IRAF package \textsc{SIRIUS} at http://optik2.mtk.nao.ac.jp/~yas/pipeline/siriuspipeline.html.}. Astrometry was done using the IRAF task \textsc{ccmap} (see Section~\ref{astrometry_sec}). The astrometrically-calibrated images were processed to remove the majority of point sources, as described in Section~\ref{starsub_sec}. Objects in the ``cleaned'' images were identified using Source Extractor (SExtractor, see Section \ref{sextr_sec}). 

The Norma cluster is located at a mean heliocentric recession velocity of 4844 $\pm$ 63 km s$^{-1}$ \citep{Woudt98}. This translates to a velocity of $v_{CMB}$ = 4928 $\pm$ 63 km s$^{-1}$ in the cosmic microwave background (CMB) rest frame \citep{Kogut93} and a distance of 70$h_{70}^{-1}$ Mpc, assuming that the cluster is at rest with respect to the CMB. Using this distance and a $k$-correction of -0.05 mag (see \citealt{Glass99}), the distance modulus for the Norma cluster becomes 
\begin{equation}\label{dist_mod_eqn}
m - M = 34.18 - 5\log{h_{70}} + A(l,b)
\end{equation}
where $A(l,b)$ is the galactic extinction.

The 47 fields cover an area of $\sim$ 0.6 square degrees, corresponding to 0.8$h_{70}^{-2}$ Mpc$^{2}$ at the distance of the cluster. The survey region is outlined by the solid line in
Fig.~\ref{obs_fig}.  It is centred on (Right Ascension (RA),
Declination (Dec)) = (16$^{\rm{h}}$15$^{\rm{m}}$01$^{\rm{s}}$.4,
-60$^\circ$54$^{\prime}$23$^{\prime\prime}$, J2000), close to the
giant elliptical galaxy at the centre of the cluster (ESO 137-G006,
WKK6269; \citealt{Woudt01}; see also \citealt{Woudt08}). The dots represent galaxies found in a deep optical search using SRC IIIaJ sky survey plates (\citealt{Woudt98}; \citealt{Woudt01}).

\subsection{Astrometry}\label{astrometry_sec}

Astrometry was done using the IRAF task \textsc{ccmap}, using stars from the 2nd generation of the Digitized Sky Survey (DSS2) images to find the astrometric solution. This was tested by comparing the positions of objects identified by SExtractor to their counterparts in the 2MASS point source catalogue (PSC; \citealt{Skrutskie06}). For 6488 point sources, the standard deviation in RA and Dec was found to be $\sigma = 0.18 \mbox{ arcsec}$. Sixty-nine extended sources were matched to objects in the 2MASS XSC, with a standard deviation of $\sigma = 0.4 \mbox{ arcsec}$.

\subsection{Star subtraction}\label{starsub_sec}

Images of the galaxies in the Norma cluster are severely affected by star-crowding, due to the low galactic latitude location of the cluster. It is necessary to clean the images of foreground stars before accurate photometry of the galaxies can be derived. \citet{Buta99} developed an IRAF routine known as \textsc{killall} to deal with this problem for the IC 342/Maffei group of galaxies. \textsc{killall} is built around the tasks in the \textsc{daophot} package \citep{Stetson87}. The \textsc{daophot psf} task builds a two-component model of the point spread function (psf), by fitting a function to a number of stars across the image. The parameters of the analytic function are found by fitting the chosen psf stars, weighted by their signal-to-noise ratios. The psf is then used to subtract stars from the image. We applied three runs of \textsc{killall} in the cleaning process, described in detail in \citet{Skelton07}. The cleaned image is interactively edited using \textsc{imedit}, to remove spikes from saturated stars, residuals from bright stars and other cosmetic defects resulting from the cleaning process. In a few cases bright stars covering galaxies are not removed by \textsc{killall}. These are dealt with individually. The galaxy is modelled using the \textsc{ellipse} and \textsc{bmodel} tasks within the \textsc{isophote} package. Sigma-clipping is applied to mask out the star light on the first iteration. The model galaxy is subtracted from the original using \textsc{imarith}. The overlapping stars are then clearly visible and can be subtracted by fitting the psf to them. In cases where the stars are saturated, this is not successful and an \textsc{imedit} aperture replacement is used. The model of the galaxy is then added back. 

\begin{figure}
	\begin{minipage}[]{4.cm}
		\resizebox{4.cm}{5.81cm}{\includegraphics{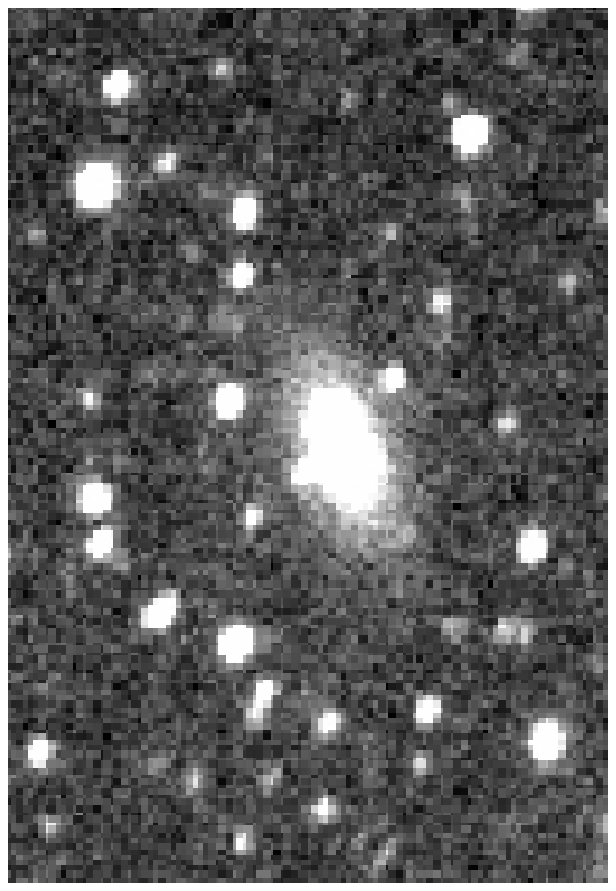}}
	\end{minipage}
	\hspace{0.2cm}
	\begin{minipage}[]{4.cm}
		\resizebox{4.cm}{5.81cm}{\includegraphics{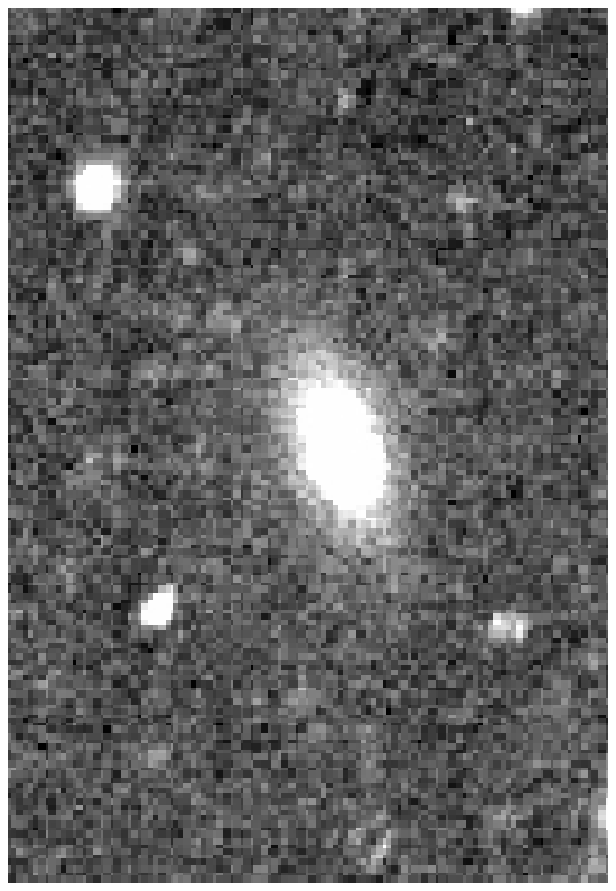}}
	\end{minipage}

\caption[An image section at the initial and final stages of the cleaning process]{\label{cleaning_pics}Examples of an image at different stages of the cleaning process. Left-hand panel: A section of the original (post data reduction) image of field 6. Right-hand panel: The same section after automated star subtraction.}
\end{figure}

A section of field 6 before and after the star removal process is shown in Fig.~\ref{cleaning_pics}. The original image after data reduction is given in the left hand panel, and the ``cleaned'' image after star subtraction in the right-hand panel.

\citet{Woudt05} tested the star removal process by adding star fields of similar density to that of the Norma cluster region to calibration galaxies with well-determined magnitudes and comparing their original photometry to that after star subtraction. They found the differences in magnitude to be negligible ($\la 0.01$~mag on average), demonstrating the effectiveness of this method for dealing with foreground stars.

\subsection{Photometry} \label{phot_sec} \label{sextr_sec}

After the star removal process, each image is run through SExtractor \citep{Bertin96}. A background mesh size of 64 pixels and filter size of 5 meshes were used for all the fields. The parameters were optimised by comparing the photometry of stars in test fields with those matched in the 2MASS PSC (see Section~\ref{photcomp_sec}). The detection threshold was chosen to be 3$\sigma$ above the background value. A gaussian convolution mask, adjusted to match the average seeing, was used to improve the detection of faint sources. 

Blended objects made up of neighbouring stars were identified by eye and used as test cases to find the best value of the deblending parameters. Increasing the parameter controlling the number of thresholds to 64 and using a minimum contrast of 0.001 improved the separation of overlapping objects. 

There are four types of photometry that can be output by SExtractor: isophotal, corrected isophotal, aperture and Kron or Auto photometry. For consistency, we have chosen to use the Auto magnitude as an estimate of the total magnitude for all the identified galaxies. 

\subsubsection{Photometric consistency checks}\label{int_check} \label{photcomp_sec}

The filter system of SIRIUS on the IRSF is approximately the same as that of 2MASS, allowing for the direct comparison of photometry. The IRSF survey of the Norma Cluster improves on the data for this region provided by 2MASS, due to the depth of the survey and differences in resolution of 2MASS and the IRSF (2.0 arcsec vs. 0.45 arcsec per pixel). When candidate galaxies are included in the sample, the completeness limit of this survey is two magnitudes fainter than the nominal completeness of 2MASS ($K_s$ = 15.$^{\rm{m}}$25 vs. $K_s$ = 13.$^{\rm{m}}$1, \citealt{Jarrett00}) and about 2.5 mag fainter than the estimated completeness at the latitude of the cluster \citep{Jarrett04}. The higher spatial resolution of the IRSF makes it possible to separate stars and extended sources to a greater extent, thus improving star subtraction for large galaxies. 

The photometry of point sources and extended sources was compared to that of corresponding objects in 2MASS, yielding a standard deviation of $\sigma_m = 0.2 \mbox{ mag}$, with no significant trend found.

Point sources identified in overlapping borders of adjacent fields were used 
to check consistency between fields. Generally of the order of 100 objects per 
field had counterparts on an adjacent field. Nine test fields were compared with the fields surrounding them, thus covering the whole region. No systematic differences in photometry over the region were found. The mean difference in magnitude and standard deviation are 
$$ <\Delta K_s> = -0.016 \pm 0.004 \mbox{ mag, }\sigma = 0.157 \mbox{ mag}.$$ 
These results reflect the maximum difference, since the objects compared are all near the edges of the images, where their parameters are least well determined. 

Some point sources were found in one field but not in an overlapping field. This was initially a cause for concern, but can be attributed to differences in image quality and the star-removal process. For example, objects towards the right-hand edge of some of the images show systematic elongation. Faint stars in affected fields appear to be extended sources. The star may be completely removed during the cleaning process on another field where the seeing is better and thus only identified on one of two overlapping fields. This was taken into account during the final selection of extended sources, where objects were examined by eye in each of the images in which they appear.


\subsection{Star-galaxy separation}

A number of techniques have been used to distinguish between stars and galaxies, usually based on two parameters, such as the magnitude and central surface brightness or magnitude and isophotal area. Objects identified by SExtractor are classified as stars or extended sources, using a Neural Network (NN) to distinguish between them. Classification is difficult for faint and blended objects. In the ZOA, the star crowding and extinction make the problem even more severe. The stellarity index given by SExtractor was unable to clearly separate stars and galaxies for objects with $K_s \ga 11$ mag, becoming unreliable for most objects beyond $K_{s} = 13$ mag. For this reason, we decided to identify galaxies by examining the images by eye. More uncertain cases are found in images with worse seeing: faint stars are blurred, becoming indistinguishable from extended sources, groups of faint stars may appear to be nebulous and even bright blended stars may appear to be a single elongated object. Bright blended stars can be distinguished from galaxies because the SB of stars drops off rapidly compared to extended sources. 

Possible extended sources identified by SExtractor were classified by examining each image by eye. The resulting catalogue was compared to the deep $B$-band catalogue from the IIIaJ Sky Survey plates (\citealt{Woudt98}; \citealt{Woudt01}). Deep uncalibrated $R_C$-band images \citep{Markus06} were used to confirm or reject objects as galaxies, or assign them ``candidate'' status if classification was still uncertain. The $R_C$-band images typically have less than 1 arcsec seeing and higher spatial resolution of 0.237 arcsec/pixel compared to the $K_s$ band image resolution of 0.453 arcsec/pixel. They are thus useful for resolving pairs or groups of stars identified as possible galaxies in the $K_s$-band. 

\section{PROPERTIES OF THE OBSERVED GALAXIES}

\label{cat.sec} \label{ext_cor.sec}

The catalogue of objects identified in the $K_s$-band contains 390 objects, 235 of which are classified as galaxies and 155 as candidate galaxies. The first fifteen entries of the catalogue are shown as an example in Table~\ref{cat_15entries.tbl}. The full catalogue and a description of the column entries are given in Appendix A. 

\begin{table*}
\scriptsize
\caption[Photometric Galaxy Catalogue]{\label{cat_15entries.tbl} The first fifteen entries of the photometric catalogue of galaxies identified in the $K_s$-band (see the electronic edition for the full catalogue)}
 \begin{tabular}{lcccccccr@{  }lcrcl}
\hline
$K_s$ ID & Optical ID & \multicolumn{2}{c}{RA (J2000)  Dec}& Gal $l$ & Gal $b$ & $A_{K_s}$ & Class & \multicolumn{2}{c} {$K_s$} & $B_J$ & Diameter & Velocity & Field/s (Flag) \\
& & [$^{h}$ $^{m}$ $^{s}$] & [$^{\circ}$ $^{\prime}$ $^{\prime\prime}$] & [$^{\circ}$]& [$^{\circ}$] & [mag] & & \multicolumn{2}{c} {[mag]} & [mag] & [$^{\prime\prime}$] & km s$^{-1}$ \\
(1) & (2) & (3a) & (3b) & (4a) &  (4b) & (5) & (6) & (7) & & (8) & (9) & (10) & (11)\\
\hline

K001 & WKK6091 & 16 11 50.8 & -60 36 14 & 325.21 & -6.72 & 0.070 & G & 13.34 $\pm$ & 0.02 & 17.9 &   8 & 3235 & 29(0)
\\ 
K002 & WKK6092 & 16 11 51.4 & -60 37 55 & 325.20 & -6.74 & 0.070 & G & 10.74 $\pm$ & 0.00 & 14.7 &  26 & 4688 & 29(2)
\\ 
K003 & WKK6090 & 16 11 52.0 & -61 11 41 & 324.81 & -7.15 & 0.084 & G & 13.17 $\pm$ & 0.02 & 16.2 &   7 & ... & 34(2)
\\ 
K004 & ... & 16 11 52.1 & -60 37 34 & 325.20 & -6.74 & 0.070 & C & 15.04 $\pm$ & 0.04 & ... &   3 & ... & 29(0)
\\ 
K005 & ... & 16 11 52.6 & -61 09 09 & 324.84 & -7.12 & 0.086 & C & 15.39 $\pm$ & 0.05 & ... &   3 & ... & 34(0)
\\ 
K006 & ... & 16 11 54.2 & -60 37 18 & 325.21 & -6.74 & 0.070 & C & 16.33 $\pm$ & 0.07 & ... &   2 & ... & 29(0)
\\ 
K007 & ... & 16 11 55.6 & -60 39 39 & 325.18 & -6.77 & 0.074 & G & 15.19 $\pm$ & 0.05 & ... &   4 & ... & 29(0)
\\ 
K008 & WKK6100 & 16 11 55.9 & -60 43 33 & 325.14 & -6.82 & 0.079 & G & 12.36 $\pm$ & 0.01 & 15.7 &   9 & ... & 30(0)
\\ 
K009 & ... & 16 11 56.8 & -61 01 05 & 324.94 & -7.03 & 0.074 & C & 15.72 $\pm$ & 0.06 & ... &   2 & ... & 32(0)
\\ 
K010 & WKK6098 & 16 11 57.1 & -61 07 53 & 324.86 & -7.11 & 0.082 & G & 14.69 $\pm$ & 0.04 & 17.1 &   4 & 10482 & 33(0) 34(0)
\\
K011 & R & 16 11 57.6 & -61 10 07 & 324.83 & -7.14 & 0.086 & G & 14.57 $\pm$ & 0.03 & ... &   4 & ... & 34(0)
\\ 
K012 & WKK6101 & 16 11 58.6 & -60 51 11 & 325.05 & -6.92 & 0.081 & G & 12.51 $\pm$ & 0.01 & 16.7 &  15 & 4416 & 31(0)
\\ 
K013 & ... & 16 11 59.4 & -60 37 43 & 325.21 & -6.75 & 0.074 & G & 13.17 $\pm$ & 0.01 & ... &  11 & ... & 29(0)
\\ 
K014 & R & 16 12 00.4 & -60 51 26 & 325.05 & -6.92 & 0.081 & G & 13.71 $\pm$ & 0.02 & ... &   7 & ... & 31(0)
\\ 
K015 & ... & 16 12 01.0 & -61 01 15 & 324.94 & -7.04 & 0.074 & C & 15.68 $\pm$ & 0.06 & ... &   3 & ... & 33(0)
\\ 
\hline
\end{tabular}
\end{table*}

\begin{figure}
	
	\resizebox{8.4cm}{7.20cm}{\includegraphics{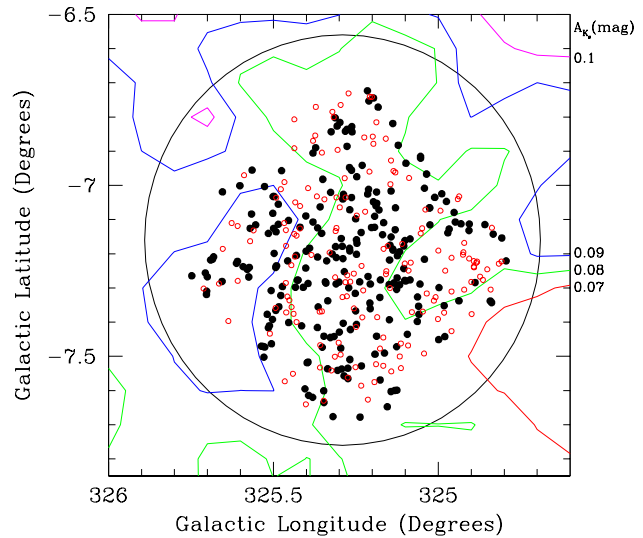}}
	
\caption[The galaxy distribution and contours of Galactic extinction]{\label{ext_map} The distribution of galaxies identified in the $K_s$-band images, in Galactic coordinates. Galaxies are marked with large filled circles, while candidates are shown as small open circles. The contours are lines of equal Galactic reddening from the dust maps of Schlegel et al. (1998), corresponding to extinction in the $K_s$-band of $A_{K_s}$ = 0.07, 0.08, 0.09 and 0.1 mag. The solid circle corresponds to $\frac{1}{3}$ Abell radius (0.6$^\circ$). 
}
\end{figure}

The distribution of objects, in Galactic coordinates, and contours of equal Galactic extinction determined from the IRAS/DIRBE maps \citep{Schlegel98} are shown in Fig.~\ref{ext_map}. The galaxies are marked with large filled circles, while candidate galaxies are shown as small open circles. The contours correspond to lines of constant extinction in the $K_s$-band, with values of $A_{K_s}$ = 0.07, 0.08, 0.09 and 0.1 mag, respectively. The extinction in the $K_s$-band is only 9\% of that in the $B$-band. It is worth noting that the small variation in mean $K_s$-band extinction over the regarded survey region implies that the contribution of extinction to uncertainties in the estimated completeness and luminosity function will be negligible, despite the location of this cluster in the ZOA.

\subsection{Sample completeness}\label{compl_lim}

\begin{figure}
 	\resizebox{8.4cm}{6.6cm}{\includegraphics{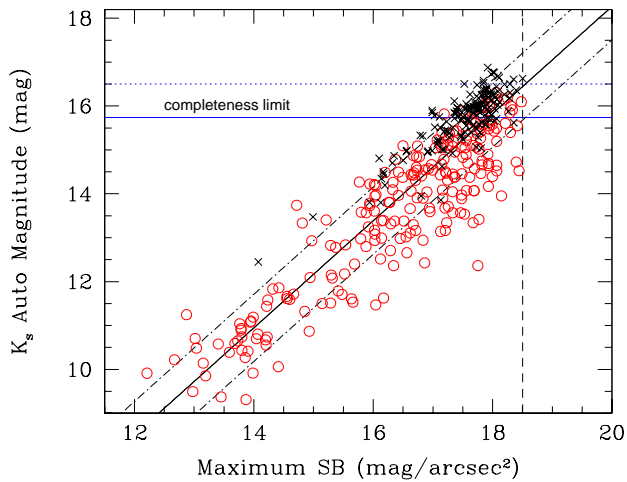}}
\caption[Surface-Brightness versus Magnitude]{\label{sb_mag.fig} Surface brightness vs. Auto magnitude. The fitted relation is indicated by the thick solid line, with the dash-dot lines representing a deviation of 1$\sigma$ on either side. The completeness magnitude is indicated by the solid line. Reliable galaxies are shown as open circles and candidates as crosses.
}

\end{figure}
The completeness limit was estimated using the method described in \citet{Garilli99} and \citet{Andreon00b}. Galaxies will not be detected when their central surface brightness (SB) drops below the detection threshold at a particular magnitude. The relationship between maximum SB, given by SExtractor, and the Auto magnitude is shown in Fig.~\ref{sb_mag.fig}. In this figure galaxies are indicated by open circles, while candidate galaxies are shown as crosses. The thick solid line is a linear fit to all the objects (+1$\sigma$ and --1$\sigma$ are shown as dash-dot lines). The dashed line indicates the SB detection limit. Galaxies to the right of the SB limit, yet below the dotted line (the intercept between the SB limit and the fitted relation) are not detected. The completeness limit is thus given by the solid line, at the magnitude where the SB limit intersects with the --1$\sigma$ relation, rather than with the dotted line.

\begin{figure}
	\resizebox{8.4cm}{5.47cm}{\includegraphics{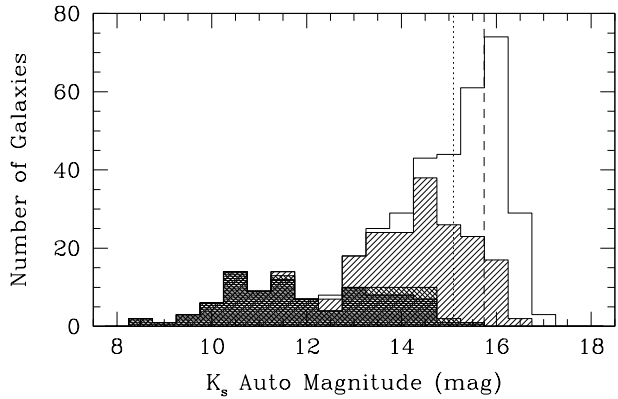}}

\caption[Number counts of galaxies with magnitude]{\label{num_counts.fig} The number counts in half-magnitude bins. The shaded region represents likely galaxies, while the unshaded region represents candidate galaxies. Galaxies with velocities less than 8000 km s$^{-1}$ are assumed to be members of the cluster, shown with the darkest shading. Non-members have the second darkest shading, galaxies with no spectroscopic information have diagonal shading and the remaining region represents candidate galaxies. The completeness limits including and excluding candidates are shown by the dashed and dotted lines, respectively.
}
\end{figure}

The completeness limit determined in this way is $K_s = 15.74$ mag. The slope of the relationship between surface brightness and magnitude changes slightly and has smaller scatter if the candidate galaxies are excluded from the fit. This results in a completeness limit of $K_s = 15.10$ mag. Fig.~\ref{num_counts.fig} shows a histogram of the number of galaxies and candidates per 0.5 mag bin. The completeness limits including and excluding candidates are indicated by the dashed and dotted lines, respectively. It is apparent from the figure that the number of reliably identified galaxies begins to drop for $K_s \ga $14.5 mag, although the total number of objects increases to $K_s \sim $ 16~mag. We adopt $K_s$~=~14.75 mag as the limiting magnitude for the determination of the LF; this sample consists of 170 galaxies and 14 candidates. 

\subsection{Comparison with optical data}\label{opt_comp}

Counterparts for 101 of the 235 (43\%) of the galaxies in our catalogue have counterparts in the catalogue obtained from a deep optical search using the IIIaJ film copies of the SRC sky survey \citep{Woudt01}. Only five galaxies identified in the optical were not detected in the $K_s$-band. The $B_J$-band magnitudes and extinctions of these galaxies are given in Table~\ref{missedB.tbl}, along with the estimated $K_s$-band magnitudes they would have if they are assumed to be late-type galaxies with a colour of $B_J - K_s \sim 2$. The brightest galaxy (WKK6369) can be seen on two of the fields, however it is partially obscured by a saturated star and thus was not detected by SExtractor. WKK6255, WKK6141 and WKK6408 were all identified by eye as low surface brightness galaxies but not extracted by SExtractor, while the faintest galaxy (WKK6410) was difficult to identify by eye. In the case of WKK6255, a comparison with the original image showed that a star was removed very close to the galaxy. These results suggest that the detection of low surface brightness galaxies may be influenced by star-subtraction and that the completeness is lower than expected from the surface brightness-magnitude relation, however these 5 galaxies amount to less than 5\% of the sample of galaxies with $B_J$-band magnitudes in the region. 

\begin{table}
\caption{\label{missedB.tbl} $B_J$-band galaxies not detected in the $K_s$-band}
\begin{tabular}{lccccc}
\hline
WKK ID & $B_J$  & $A_{B}$ & $A_K$ & Approx. $K_s$\\
& [mag] & [mag] & [mag] & [mag]\\
\hline
WKK6369$^{\footnotesize{a}}$ & 16.5	& 0.855	&  0.075	& 13.7\\
WKK6255 & 17.2	& 0.976	&  0.086	& 14.3\\
WKK6141 & 17.4	& 0.829	&  0.073	& 14.6\\
WKK6408 & 18.4	& 1.04	&  0.091	& 15.5\\
WKK6410 & 18.9	& 1.04	&  0.092	& 16.0\\
\hline
\end{tabular}
\footnotesize{$^{a}$Covered by a saturated star in the $K_s$-band image}
\end{table}

An additional 89 galaxies (38\%) were previously identified on deep $R_C$-band images of the Norma cluster taken with the Wide Field Imager on the MPG/ESO 2.2-m telescope. Velocity information is available for 102 galaxies (43\% of the total galaxy sample); these velocities come from a variety of observations following our Zone of Avoidance redshift surveys (SAAO: \citealt{Woudt99}, MEFOS: \citealt{Woudt04}, Parkes: Schroeder, Kraan-Korteweg \& Henning, in preparation) complemented by additional pointed observations of the Norma cluster members with 2dF (see \citealt{Woudt08} for details regarding the optical and near-infrared target selection of the 2dF observations). Galaxies with velocities greater than 8000~km~s$^{-1}$ ($N=9$) are assumed to be background galaxies. The mean velocity of cluster galaxies in the survey area with velocity information, excluding the ones with $v > 8000$~km~s$^{-1}$, is 4816 km~s$^{-1}$ with a dispersion of 978~km~s$^{-1}$. This agrees well with the value derived in paper I on the dynamical analysis of this cluster for the inner 1/3$R_A$, which was found to be $4777 \pm 86$~km~s$^{-1}$, $\sigma = 973$~km~s$^{-1}$ for 129 cluster members.

\section{THE $K_s$-BAND LUMINOSITY FUNCTION}

\subsection{Derivation of the luminosity function}

The magnitudes of each galaxy are corrected for extinction using the colour excess given by the IRAS/DIRBE maps \citep{Schlegel98}, converted to a magnitude correction using the NIR extinction law of \citet{Cardelli89}. No correction is made for stellar confusion because the majority of stars have already been removed during the star subtraction process.  The number of objects that are misclassified as galaxies is also reduced by classifying all the objects by eye. Uncertain cases are labelled as candidates, thus the LF determined for certain galaxies can be seen as a lower limit. 

The largest source of uncertainty in the determination of the LF is the correction per magnitude bin for background contamination. This is often done statistically through a control field away from the cluster, though this can lead to signficant systematic errors (see e.g. \citealt{Andreon05}, for a detailed discussion). However, background contamination can also be assessed when membership assignment is available for a (fair) fraction of the cluster sample. In our case, spectroscopic information is available for 99 of the 170 $K_s$-band galaxies brighter than the completenesss limit (58\%). Although our number counts are lower, we will follow the method used by \citet{Mobasher03} to correct a large spectroscopic survey in the Coma cluster for incompleteness. This approach is similar to that of \citet{Biviano95}, who make a statistical estimate of the membership fraction to confirm an estimate based on selection in the colour-magnitude plane. There is a potential bias introduced by this method due to the selection being based on a spectroscopic sample which might favour bluer galaxies. It should be noted though that in this case the spectroscopically observed galaxies were selected from both optical and NIR samples. Moreover neither aimed for a given magnitude completeness limit, hence a 'systematic' colour bias is not expected. For reassurance in this regard, we compare the LF obtained with this spectroscopic method with that obtained by subtracting $K$-band galaxy counts of field samples. As is shown later, these vary only for the faintest bins and have no influence on the LF parameters, which are stable for the last three magnitude bins used for the fitting of the LF. We only include galaxies with $K_s \lt 14.75$ mag in the discussions of the LF henceforth, because the membership fraction beyond this is not reliably determined (only two galaxies have redshifts, one of which is a member of the cluster).

\subsubsection{Cluster membership}

Let the number of galaxies in the area covered by the $K_s$-band survey with spectroscopic information be $N_s(m)$, in a given range of apparent magnitude. None of the galaxies in the spectroscopic sample have velocity $\la$ 2000 km s$^{-1}$, which would place them as foreground galaxies. All galaxies with velocities less than 8000 km s$^{-1}$ are thus assumed to be members of the cluster. Of the 99 galaxies with velocities, 8 (8\%) are background. Let the number of galaxies that are members and non-members of the cluster as a function of magnitude be $N_m(m)$ and $N_{nm}(m)$ respectively. The fraction of galaxies satisfying the membership criterion in each magnitude bin is then $$ f(m) = \frac{N_m(m)}{N_m(m) + N_{nm}(m)} = \frac{N_m(m)}{N_s(m)}.$$ The membership fraction, presented in Table~\ref{frac_mem.tbl}, gives the probability that a random galaxy drawn from the spectroscopic sample will be a member of the cluster, as a function of magnitude. The number of members is a binomial random variable, with the probability of ``success'' given by the true membership fraction for each of the $N_s$ trials. The error in the membership fraction, shown in brackets in Table~\ref{frac_mem.tbl}, is thus given by 
\begin{eqnarray}\label{err_f}
	\mbox{var}[f(m)]^{1/2} & = & \frac{\mbox{var}[N_m(m)]^{1/2}}{N_s(m)} \nonumber \\
		  & = & \left[\frac{f(m)(1-f(m))}{N_s(m)}\right]^{1/2} 
\end{eqnarray}

\begin{table}
\caption{\label{frac_mem.tbl} Spectroscopic number counts and the membership fraction in $K_s$-band magnitude intervals}
\begin{minipage}{8.5cm}
\begin{center}
\begin{tabular}{lrrr}
\hline
\multicolumn{1}{c}{$K_s^0$ [mag]} & \multicolumn{1}{c}{$N_s(m)$} & \multicolumn{1}{c}{$N_{m}(m)$} & \multicolumn{1}{c}{$f(m)$} \\
\hline
8.50 & 2.00 & 2.00 & 1.00 (0.00) \\ 
9.00 & 2.00 & 2.00 & 1.00 (0.00) \\ 
9.50 & 2.00 & 2.00 & 1.00 (0.00) \\ 
10.00 & 7.00 & 7.00 & 1.00 (0.00) \\ 
10.50 & 14.00 & 14.00 & 1.00 (0.00) \\ 
11.00 & 9.00 & 9.00 & 1.00 (0.00) \\ 
11.50 & 13.00 & 12.00 & 0.92 (0.07) \\ 
12.00 & 6.00 & 6.00 & 1.00 (0.00) \\ 
12.50 & 6.00 & 6.00 & 1.00 (0.00) \\ 
13.00 & 8.00 & 8.00 & 1.00 (0.00) \\ 
13.50 & 11.00 & 9.00 & 0.82 (0.12) \\ 
14.00 & 9.00 & 7.00 & 0.78 (0.14) \\ 
14.50 & 10.00 & 7.00 & 0.70 (0.14) \\ 
\hline
\multicolumn{4}{c}{{\footnotesize{\textsc{NOTE.--}Errors for the membership fraction}}}\\
\multicolumn{4}{c}{{\footnotesize{are given in parentheses.}}}\\
\end{tabular}
\end{center}
\end{minipage}

\end{table}

\subsubsection{The luminosity function}

The number counts in the $K_s$-band photometric sample ($N_p(m)$) and spectroscopic sample ($N_s(m)$) per magnitude bin are shown in Fig.~\ref{num_counts.fig}. The darkest shading refers to members of the cluster, as defined above, the cross-hatched area to non-members, the diagonal shading to galaxies without spectroscopic information and the open histogram the total number of objects in the photometric sample, including candidate galaxies. If the spectroscopic sample reflects the proportions of members and non-members that can be expected in the full photometric sample, the number of galaxies without velocity information that are cluster members can be estimated as $f(m) N_{nv}(m)$, where $N_{nv}(m)$ is the number of galaxies without velocity information, in a given range of apparent magnitude.

\begin{table*}
\caption{\label{num_counts.tbl} The total number counts and resultant LF}
\begin{minipage}{126mm}
{\small{

\begin{tabular}{rrrrrrr}
\hline
& \multicolumn{3}{c} {All objects} & \multicolumn{3}{c} {Excluding candidates}\\
\hline
\multicolumn{1}{c}{$K_s$}  & \multicolumn{1}{c}{$N_p(m)$} & \multicolumn{1}{c}{$N_p(m)f(m)$} & \multicolumn{1}{c}{$\phi(m)$} & \multicolumn{1}{c}{$N_p(m)$} & \multicolumn{1}{c}{$N_p(m)f(m)$} & \multicolumn{1}{c}{$\phi(m)$}\\
\multicolumn{1}{c}{$[\rm{mag}]$} & & & \multicolumn{1}{c}{[$\rm{mag^{-1} Mpc^{-2}}$]} & & & \multicolumn{1}{c}{[$\rm{mag^{-1} Mpc^{-2}}$]}\\
\hline
8.50....... & 2.00 & 2.00 & 5.00 (3.54) & 2.00 & 2.00 & 5.00 (3.54)\\
9.00....... & 2.00 & 2.00 & 5.00 (3.54) & 2.00 & 2.00 & 5.00 (3.54)\\
9.50....... & 2.00 & 2.00 & 5.00 (3.54) & 2.00 & 2.00 & 5.00 (3.54)\\
10.00...... & 7.00 & 7.00 & 17.50 (6.61) & 7.00 & 7.00 & 17.50 (6.61)\\
10.50...... & 14.00 & 14.00 & 35.00 (9.35) & 14.00 & 14.00 & 35.00 (9.35)\\
11.00...... & 9.00 & 9.00 & 22.50 (7.50) & 9.00 & 9.00 & 22.50 (7.50)\\
11.50...... & 14.00 & 12.92 & 32.31 (9.01) & 14.00 & 12.92 & 32.31 (9.01)\\
12.00...... & 6.00 & 6.00 & 15.00 (6.12) & 6.00 & 6.00 & 15.00 (6.12)\\
12.50...... & 10.00 & 10.00 & 25.00 (7.91) & 9.00 & 9.00 & 22.50 (7.50)\\
13.00...... & 19.00 & 19.00 & 47.50 (10.90) & 19.00 & 19.00 & 47.50 (10.90)\\
13.50...... & 27.00 & 22.09 & 55.23 (13.21) & 24.00 & 19.64 & 49.09 (12.21)\\
14.00...... & 28.00 & 21.78 & 54.44 (14.14) & 25.00 & 19.44 & 48.61 (13.02)\\
14.50...... & 44.00 & 30.80 & 77.00 (19.72) & 37.00 & 25.90 & 64.75 (17.12)\\
\hline
\multicolumn{7}{c}{{\footnotesize{\textsc{NOTE.--}Errors for the LF are given in parentheses.}}}

\end{tabular}
}}
\end{minipage}
\end{table*}

The LF can then be estimated as 
\begin{eqnarray}
	\phi (m) & = & \frac{1}{A}(N_m(m) + f(m) N_{nv}(m)) \nonumber \\
		& = & \frac{N_p(m)}{A}f(m) ,
\end{eqnarray}
where $A$ is the survey area \citep{Mobasher03}. The relative error in the LF is given by the quadrature sum of the relative errors in $N_p(m)$ and $f(m)$, so 
\begin{equation}
	\frac{\mbox{var}[\phi(m)]}{\phi(m)^2} = \frac{\mbox{var}[N_p(m)]}{N_p(m)^2} + \frac{\mbox{var}[f(m)]}{f(m)^2}.
\end{equation}
The number of galaxies in the photometric sample is a Poisson variable, which has variance equal to its expectation value, while the variance of the membership fraction is given by Equation \ref{err_f}. This results in a relative error of
\begin{equation}\label{LFerr}
	\frac{\delta\phi(m)}{\phi(m)} = \left[\frac{1}{N_p(m)} + \frac{1}{N_m(m)} - \frac{1}{N_s(m)}\right]^{1/2}.
\end{equation}
Equation \ref{LFerr} reduces to $1/\sqrt{N_m(m)}$ if all the galaxies in the photometric sample at magnitude $m$ have spectroscopic data and $1/\sqrt{N_p(m)}$ if all the galaxies in the spectroscopic sample at magnitude $m$ are members of the cluster. 

The LFs determined from the membership-corrected galaxy counts are plotted in the top panel of Fig.~\ref{LF.fig}. The black histogram shows the raw counts, while the circular data points represent the LF obtained applying the above method to the total number counts (left-hand column of Table~\ref{num_counts.tbl}) and the triangular data points to certain galaxies only (candidates excluded), as given in the right-hand column of Table~\ref{num_counts.tbl}. The two LFs are very similar over the entire magnitude range considered.

A correction for background contamination can also be made by subtracting an estimate of the field LF from the raw counts found in the cluster region. In the top panel of Fig.~\ref{LF.fig} we show the field counts from \citet{Glazebrook94} as a blue dashed line and the field-subtracted LF with blue squares. Errors are assumed to be Poissonian. The counts given by \citet{Huang97} are very similar to those of \citet{Glazebrook94} for $K_s \le 14$ mag, but slightly higher for fainter galaxies and would thus result in larger corrections at the faint end. The observations of \citet{Glazebrook94} are closer to ours in terms of spatial resolution, however the area covered by the survey is much smaller than that of \citet{Huang97}. As an additional check, we subtracted the counts from \citet{Cluver08}, which were obtained using the same instrument, observing strategy and exposure times as our cluster survey, over a field of 2.24 deg$^2$. These counts are slightly higher than the other field estimates for $K_s \lt 14$ mag, possibly due to a slight overdensity at higher redshifts (12000 - 15000 km~s$^{-1}$) or due to an overestimate of the extinction. These data independently confirm our completeness limit, dropping below the other field estimates for $K_s \gt 14.75$ mag. All three field-subtracted LFs agree well with the membership fraction-corrected estimate, deviating slightly for $K_s \gt 13.75$ mag.

\subsection{The shape of the luminosity function}

\begin{figure}
	\resizebox{8.4cm}{7.46cm}{\includegraphics{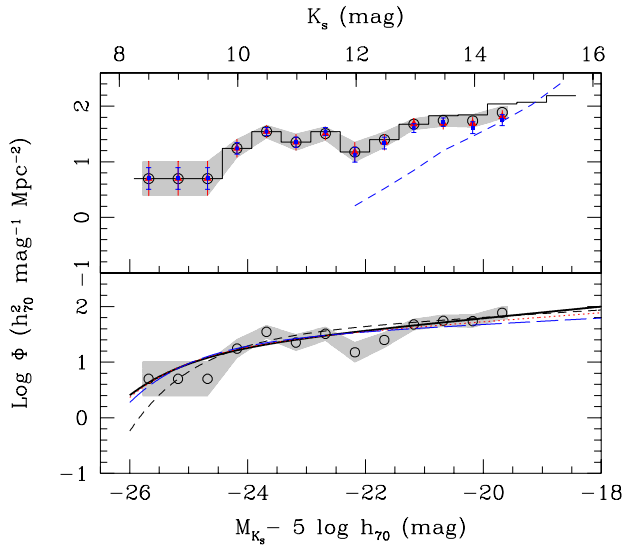}}

\caption[The $K_s$-band LF of the Norma cluster with app and abs magnitude]{\label{LF.fig} The LF of the Norma cluster in the $K_s$-band (see text for details). Top panel: The black histogram shows the raw number counts. Open black circles and grey shading represent the membership fraction-corrected LF including all objects and the corresponding error, while red triangles exclude candidate galaxies. The blue squares show the field-subtracted LF, using the \citet{Glazebrook94} estimate of the field number counts (shown as the blue dashed line). Lower panel: Open black circles and grey shading are as in the top panel. The thick solid black line is the fitted Schechter function for all objects, while the black dashed line excludes the dip point and brightest bin. The red dotted line is the fitted Schechter function excluding candidate galaxies and the blue long-dash line is the fit to the field-subtracted LF.
}
\end{figure}

The LF, shown in Fig.~\ref{LF.fig}, is flat for the three brightest bins, increasing thereafter. There appears to be a shallow dip at $K_s = 12$~mag, corresponding to an absolute magnitude of $M_{K_s} = -22.18$~mag (using a distance modulus of $34.18$~mag, see Section~\ref{obs_sec}). 

The LF can be fitted with a Schechter function \cite{Schechter76} of the form
\begin{equation}\label{schech_fn}
\phi(M) dM = \phi^* 10^{0.4(\alpha + 1)(M^* - M)}e^{{-10^{0.4(M^* - M)}}} dM
\end{equation}
The characteristic magnitude, $M^*$, faint-end slope, $\alpha$, and normalisation, $\phi^*$, were found using a nonlinear least-squares Marquardt-Levenberg algorithm. The resulting parameters are shown in Table~\ref{LF_results} and the Schechter functions including and excluding candidates are plotted in the lower panel of Fig.~\ref{LF.fig} (solid and dotted lines, respectively).

\begin{table*}
\caption{\label{LF_results} Resulting Schechter parameters from fits to the observed LF}
\begin{minipage}{126mm}
\begin{center}
{\small{
\begin{tabular}{lccrcrc}
\hline
Data fitted  & $M_{K_s}^* - 5\log h_{70}$ & $\alpha$ & \multicolumn{1}{c}{$\phi^*$} & $\chi^2$ & \multicolumn{1}{c}{$\nu$} & P($>\chi^2|\nu$)\\
             & [mag]             & & [$\rm{h_{70}^{2} Mpc^{-2}}$] & & & [\%]\\
\hline
Including candidates & & & & & &\\

All bins........................... & -25.39 $\pm$ 0.80 & -1.26 $\pm$ 0.10 & 17.06 $\pm$ 1.79 & 13 & 10 & 22\\ 
Exclude $K \gt 13.^{\rm{m}}75$....... & -25.18 $\pm$ 0.81 & -1.20 $\pm$ 0.16 & 21.26 $\pm$ 1.93 & 12 & 8 & 14\\ 
Exclude $K \gt 14.^{\rm{m}}25$....... & -25.21 $\pm$ 0.74 & -1.21 $\pm$ 0.13 & 20.55 $\pm$ 1.78 & 12 & 9 & 20\\ 
Dip point flagged............ & -25.04 $\pm$ 0.44 & -1.21 $\pm$ 0.08 & 24.01 $\pm$ 1.44 & 5.6 & 8 & 69\\ 
Exclude $K = 8.^{\rm{m}}5$.......... & -25.09 $\pm$ 1.07 & -1.23 $\pm$ 0.13 & 20.08 $\pm$ 2.05 & 13 & 9 & 17\\ 
Exclude $K = 8.^{\rm{m}}5$ \& dip & -24.54 $\pm$ 0.47 & -1.16 $\pm$ 0.09 & 32.88 $\pm$ 1.47 & 4.8 & 7 & 69\\ 
\hline
Excluding candidates & & & & & &\\

All bins.......................... & -25.23 $\pm$ 0.70 & -1.20 $\pm$ 0.10 & 20.21 $\pm$ 1.67 & 13 & 10 & 24\\ 
\hline
Field-subtracted & & & & & &\\

All bins.......................... & -25.02 $\pm$ 0.57 & -1.14 $\pm$ 0.11 & 25.39 $\pm$ 1.57 &  30 & 10 & 0.10\\ 
\hline

\end{tabular}
}}
\end{center}

\end{minipage}
\end{table*}

The effects of removing one or more data points from the fit are shown in Table~\ref{LF_results}. The Schechter parameters were found to be robust to changes in the limiting magnitude. Similar values (within $1\sigma$ of the fit to all the data) were obtained for all three parameters, using limiting values from $K_s = 13.75$ mag (excluding the last 2 bins) to $K_s = 14.75$ mag (including all bins, the solid line in Fig.~\ref{LF.fig}). In some clusters, the fit to a Schechter function is improved by excluding the brightest cD galaxy from the fit \citep{Binggeli88}. In this case, excluding the brightest bin, which contains the two cD galaxies, produces a slightly fainter characteristic magnitude but does not significantly change the results. The fit improved slightly by flagging the dip point (bins $K_s = 12$ mag and $K_s = 12.5$ mag), however the characteristic magnitude and faint end slope remain within $1\sigma$ of the parameters obtained by fitting all the data. The uncertainty on the parameters is approximately halved. The most dramatic change occurs when both the dip point and the brightest bin are excluded. The normalisation increases substantially and the characteristic magnitude is fainter by more than $1\sigma$.  This fit is shown as the dashed line in Fig.~\ref{LF.fig}. The Schechter function fit to the field-subtracted LF (long dashed line in Fig.~\ref{LF.fig}) agrees well with the other fits. The $\chi^2$ parameter, number of degrees of freedom, $\nu$, and probability of obtaining a larger $\chi^2$, P($>\chi^2|\nu$), for each fit are listed in Table~\ref{LF_results}. 

Various statistical tests into the reality of the dip point were performed, including a set of Monte Carlo simulations selecting 10000 random Schechter functions from within the errors of the fitted Schechter function parameters and comparing the resultant distribution with the data. In addition, we used the Bayesian and Aike information criteria to assess whether a more complicated model (a Gaussian at the bright end, with a Schechter, second Gaussian or power-law at the faint end) provides a better fit to the LF \citep[e.g][]{Liddle07}. Despite the coincidence of the dip point in our LF with those seen in other LFs at $M_{K_s} = -22.18$~mag, the statistical tests show that the dip in the Norma LF, based on our present data, is not significant. The reality of the dip in the Norma LF can only be ascertained by enlarging the areal coverage and thereby reducing the errors in each magnitude bin. 

\subsection{Comparison to other luminosity functions}
\label{coma_comp}

The derived LF of the Norma cluster in the $K_s$-band is comparable in form to the LFs of other clusters and the field LF. The comparison of the Schechter parameters is not straight forward, however, due to differences in wavelength band, magnitude system, area covered by the survey, depth of the survey, region within the cluster etc. In this section, we compare our LF to that of the Coma cluster, an intermediate redshift cluster, AC118, and the field. All magnitudes from the literature used in the comparison have been converted to H$_0$ = 70$h_{70}$~km s$^{-1}$ Mpc$^{-1}$.

The characteristic magnitude for the Coma cluster found by \citet{DePropris98} in the $K$-band ($M^*_K = -24.^{\rm{m}}02$) and \citet{Andreon00a} in the $H$-band ($M^*_H = -23.86$~mag) agree well, taking into account the colour difference $H - K \approx 0.22$~mag \citep{DePropris98}.  No errors for $M^*$ are quoted in \citet{DePropris98}, however the $1\sigma$ error ellipse shown in their Fig.~2 indicates that $1\sigma \sim 0.75$~mag. The confidence interval for the LF fit of \citet{Andreon00a} is even larger (their Fig.~3). The characteristic magnitude of $M^*_{K_s} = -25.39\pm 0.80$~mag for the Norma cluster is brighter than for Coma, comparing the parameters at face value. Excluding the dip point and brightest bin, the characteristic magnitude of the Norma cluster is $M^*_{K_s} = -24.54 \pm 0.47$~mag, within $1\sigma$ of these Coma LF determinations. 
\begin{table*}
\begin{center}

\begin{minipage}{126mm}
\begin{center}
\caption{\label{fix_alpha_LF.tbl} Resulting Schechter parameters from fits to the observed LF, with various adopted fixed faint end slopes}
\begin{tabular}{ccrcr}
\hline
& \multicolumn{2}{c}{All Bins} & \multicolumn{2}{c}{``Best Fit''$^{\footnotesize{a}}$}\\
\hline
$\alpha$ & $M_{K_s}^* - 5\log h_{70}$ & \multicolumn{1}{c}{$\phi^*$} & $M_{K_s}^* - 5\log h_{70}$ & \multicolumn{1}{c}{$\phi^*$}\\
             & [mag] & \multicolumn{1}{c}{[$\rm{h_{70}^{2} Mpc^{-2}}$]} & [mag] & \multicolumn{1}{c}{[$\rm{h_{70}^{2} Mpc^{-2}}$]}\\
\hline
 -0.98 & -24.41 $\pm$ 0.27 & 48.83 $\pm$ 1.15 & -23.99 $\pm$ 0.18 & 59.00 $\pm$ 1.11\\ 
 -1.00 & -24.46 $\pm$ 0.27 & 46.01 $\pm$ 1.15 & -24.04 $\pm$ 0.19 & 55.70 $\pm$ 1.11\\ 
 -1.16 & -24.94 $\pm$ 0.35 & 26.42 $\pm$ 1.15 & -24.56 $\pm$ 0.25 & 32.28 $\pm$ 1.12\\ 
 -1.20 & -25.11 $\pm$ 0.40 & 22.31 $\pm$ 1.17 & -24.74 $\pm$ 0.30 & 27.24 $\pm$ 1.14\\ 
 -1.30 & -25.68 $\pm$ 0.68 & 13.25 $\pm$ 1.29 & -25.41 $\pm$ 0.64 & 15.88 $\pm$ 1.28\\ 
\hline
\multicolumn{5}{l}{\footnotesize{$^{a}$Excludes $K = 8.^{\rm{m}}5$ \& dip}}

\end{tabular}
\end{center}
\end{minipage}
\end{center}
\end{table*}

We obtain an intermediate value of $\alpha$, ranging from -1.14 to -1.26. This is steeper than the \citet{DePropris98} Schechter function fit, but shallower than the power-law slope they fit to the faint end, as well as the values of $\alpha$ obtained by \citet{Mobasher98} and \citet{Andreon00a}. A number of authors have found a steep faint-end slope for Coma, in the optical (e.g. \citealt{Secker96}, \citealt{Trentham98}) and NIR (e.g. \citealt{Mobasher98}, \citealt{DePropris98}). Most of these studies use statistical background subtraction and extend to fainter magnitudes, thus an upturn in the LF may only occur for fainter magnitudes than considered here. De Propris et al. (1998) find an upturn in the slope for $M^{*} + 3$~mag, well within the limits of this survey, however. It is possible that a Gaussian or Schechter function fitted to the bright end with a power-law fitted to the faint end is an equally valid assumption of the form of the LF. There has also been a suggestion that the slope in the central region of clusters may be shallower than in the outskirts (e.g. \citealt{Lobo97}), with evidence for this in Coma \citep{Andreon00a} and AC118 \citep{Andreon01}. A similar analysis using fields offset from the center would be necessary to confirm whether this is the case for the Norma cluster.

There is some degeneracy between the Schechter parameters, thus a further comparison can be made by fixing the faint-end slope and re-fitting $M^*$ and $\phi^*$. The results for various fixed values of $\alpha$ are shown in Table~\ref{fix_alpha_LF.tbl} for fits to all the data and the ``best fit'' case, where the brightest bin and dip points are excluded. The values of $\alpha$ have been chosen to match a selection of those in the literature. The results indicate that $M^*$ is fainter and $\phi^*$ higher for flatter faint-end slopes. For $\alpha = -0.98$, corresponding to the slope measured by \citet{DePropris98} for the bright end of the LF, $M^*$ agrees well with their characteristic magnitude. For all choices of $\alpha > -1.20$, $M^*$ is within $1\sigma$ of the \citet{DePropris98} and \citet{Andreon00a} values, in both the fits. A steeper faint slope of $\alpha = -1.30$, however, produces an even brighter $M^*$ which does not agree with previous determinations. The parameters are not well constrained for more negative values of $\alpha$. 

In order to explore the evolution of the LF, we compare the LF of the Norma cluster to that of AC118, a cluster at $z = 0.3$. The NIR LF of AC118 has been determined by \citet{Barger96} and \citet{Andreon01}, with the latter study being one of the deepest determinations of an intermediate redshift cluster LF placing constraints on both $\alpha$ and $M^*$. We note that the $K_s$-band corresponds approximately to the rest-frame $H$-band at $z = 0.3$. The transformation is $$ M_{H_{z=0}} = M_{K_{z = 0.3}} + 0.09 \mbox{ mag} $$ \citep{Andreon01}. An offset of $\sim 0.22$ mag, corresponding to the $H - K$ colour, gives the relation $$ M_{K_{z = 0}} \sim M_{K_{z = 0.3}} - 0.13 \mbox{ mag}.$$ A fixed slope of -1.0 yields a characteristic magnitude in agreement with that of \citet{Barger96} ($M^*_{K'} = -24.5$ mag), while a slope of -1.2 gives a brighter $M^*$, corresponding to $M^*_{K_s} = -25.26$ mag measured by \citet{Andreon01}. This suggests that there is little evolution in the LF from intermediate redshifts to today.

A recent determination of the field LF \citep{Heath Jones06}, using the 6dFGS, finds that the Schechter function is not an ideal fit to the LF. The Schechter function is unable to reproduce the flat faint-end slope and drop-off at the bright end simultaneously. Nevertheless, a Schechter function provides a rough approximation to the LF. The Schechter parameters obtained are $M^*_{K_s} = -24.60 \pm 0.03$ mag and $\alpha = -1.16 \pm 0.04$. These results are within $1\sigma$ of the fit to all the data and very similar to the fit with the dip and brightest bin flagged, given in Table~\ref{LF_results}, where all the parameters are free to vary. Even better agreement between the characteristic magnitudes are obtained with a fixed slope of -1.16, as shown in Table~\ref{fix_alpha_LF.tbl}. No correction has been made to account for the different types of magnitudes used, however (\citealt{Heath Jones06} subtract 0.135 mag from $M^*$ of previous 2MASS LFs to convert from Kron to total magnitude). We also note that 2MASS is known to miss flux from the outer regions of galaxies, as well as low surface brightness galaxies that are bright enough to be included in the sample, which may affect the shape of the LF (see for e.g. \citealt{Andreon02}, \citealt{McIntosh06}). It is somewhat surprising that the agreement with the field LF is so strong, given that the environments are so different. The errors on the cluster LF parameters are much larger than for the field, due to the large number of galaxies used in the field LF determination, and thus the favourable comparison may be fortuitous, rather than supportive of a universal LF. 

\subsection{Mass-to-light estimate}

An estimate of the mass-to-light ratio of the Norma cluster was derived 
using the dynamical mass of the cluster (see also \citealt{Woudt08}) and 
the total $K_s$-band luminosity of the cluster, both within the area 
surveyed here. The dynamical mass is $M_{\rm RVT} = 5.2 \pm 0.3 \times 
10^{14}$ M$_{\odot}$ and the total $K_s$-band luminosity is $L_{K_s} = 5.0 
\times 10^{12}$ L$_{\odot}$. The latter includes the 155 candidate 
galaxies, which only make a minor contribution ($0.1 \times 10^{12}$ 
L$_{\odot}$) to the total luminosity of the Norma cluster. This results 
in a mass-to-light ratio of $M/L_{K_s} = 104 \pm 6$ 
M$_{\odot}$/L$_{\odot}$, which is comparable to the mass-to-light ratio 
of the Coma cluster (\citealt{Rines01}, 2004) at similar cluster-centric 
radii. No attempt has been made to correct for flux coming from galaxies outside of the field covered in this work.

\section{SUMMARY OF RESULTS}

The $K_s$-band LF of the central 0.8$h_{70}^{-2}$ Mpc$^2$ of the Norma cluster has been determined for $M_{K_s} \la -19$ mag. The Norma cluster is a rich, nearby cluster, comparable to the Coma cluster, yet it has received little attention due to its position near the galactic plane. It has been shown that semi-automated star-subtraction techniques can be effectively used on NIR ZOA data, enabling accurate photometry of galaxies at low latitudes. IRAF routines, encompassing the \textsc{KILLALL} routine of \citet{Buta99}, were developed and implemented to remove contaminating star-light. The effects of foreground dust were taken into account using the dust maps of \citet{Schlegel98}. The detection and photometry of galaxies was done using SExtractor \citep{Bertin96}, producing a catalogue containing 390 objects, 235 of which are classified as likely or definite galaxies and 155 as candidate galaxies. The full catalogue is presented in the electronic edition.  

The $K_s$-band photometry was used to construct the LF of the cluster. The number of background galaxies at each magnitude was estimated from the membership fraction of a smaller spectroscopic sample, as well as by subtracting field counts from the literature. The results are summarised as follows:
\begin{enumerate}
	\item The constructed LF has a similar shape to the NIR LF of other clusters and the field. A Schechter function with a characteristic magnitude of $M^*_{K_s}$ = -25.39 $\pm$ 0.80 mag and faint-end slope of $\alpha$ = -1.26 $\pm$ 0.10 can be fit to the data. 

	\item We obtain an intermediate faint-end slope, which falls in the mid-range of previous determinations of the NIR LF for the Coma cluster (\citealt{Andreon00a}, \citealt{DePropris98}, \citealt{Mobasher98}). Similar intermediate slopes have been found in optical LFs of Coma \citep{Mobasher03} and Centaurus (\citealt{Jerjen97}, \citealt{Chiboucas06}) and NIR LFs of AC118 (\citealt{Barger96}, \citealt{Andreon01}). The limiting magnitude is $M_{K_s} \sim -19.5$ mag, thus the LF extends to $M^*_{K_s} + 6$, within the dwarf regime.

	\item $M^*_{K_s}$ is brighter than found in previous determinations of the NIR LF of the Coma cluster (\citealt{DePropris98}, \citealt{Andreon00a}), which may indicate that there is a higher proportion of bright galaxies in the Norma cluster. Fixing $\alpha$ to the value of \citet{DePropris98} brings the characteristic magnitudes within $1\sigma$ of each other, however. 

	\item The fit to a Schechter function is improved by excluding the brightest bin and dip, though this lacks physical motivation. The resultant parameters are $M^*_{K_s} = -24.54 \pm 0.47$ and $\alpha = -1.16 \pm 0.09$. This characteristic magnitude is compatible with those of Coma found by \citet{DePropris98} and \citet{Andreon00a}.

	\item Both the fit to all the data and the improved fit agree with the Schechter parameters found for the field from the 6dFGS \citep{Heath Jones06}. 

\end{enumerate}

In this study, the $K_s$-band LF of the Norma cluster has been determined  to the depth of dwarf galaxies over a relatively large area. This provides a benchmark of the LF of a local rich cluster, which can be compared to the Coma cluster and more distant clusters for evolutionary studies. The error bars on the LF are still large however, and future work will aim to provide better constraints on the LF parameters. To work toward this, we increased the survey area. A comparison of the $K_s$-band LF of the central region from this study, with outer regions or a larger area encompassing the centre, will then be used to investigate trends such as the increase in the faint-end slope of the LF with clustercentric radius and other environmental effects. We also plan to analyse the $J$- and $H$-bands, as well as the independently obtained $R_C$-band survey. This would provide insight into the complex make-up of the cluster and give a better understanding of the shape of the total LF and significance of the dip. 

\section*{ACKNOWLEDGMENTS}

We would like to thank the referee, S. Andreon, for a very thorough and helpful report, which greatly improved the paper. Financial support by the South African National Research Foundation is gratefully acknowledged. RES is a member of the International Max Planck Research School for Astronomy and Cosmic Physics and would like to thank Eric Bell for useful discussions and support. This research has made use of the NASA/IPAC Infrared Science Archive (2MASS) and the NASA/IPAC Extragalactic Database (NED), which are operated by the Jet Propulsion Laboratory, California Institute of Technology, under contract with the National Aeronautics and Space Administration.

\appendix
\section{Photometric catalogue}

The catalogue of objects identified in the $K_s$-band contains 390 objects, 235 of which are classified as galaxies and 155 as candidate galaxies, as described in Section \ref{cat.sec}. The photometric properties of these galaxies are given in Table \ref{cat.tbl}. The table contains the following columns:\\
\textbf{Column 1}: Identification number from this survey. \\
\textbf{Column 2}: Identification number from the catalogue of Woudt \& Kraan-Korteweg (2001), if a corresponding object was found in their deep optical search of the SRC IIIaJ sky survey plates, or an `R' indicating identification on the $R$-band images of the cluster.\\
\textbf{Column 3}: Right Ascension (RA) and Declination (Dec) (J2000)\\
\textbf{Column 4}: Galactic longitude ($l$) and latitude ($b$) in degrees\\
\textbf{Column 5}: Galactic Extinction in the $K_s$-band ($A_{K_s}$) in magnitudes, as derived from the Schlegel, Finkbeiner \& Davis (1998) reddening maps.\\
\textbf{Column 6}: Classification as likely galaxy (G) or candidate (C), based on eye-balling of the $K_s$- and $R$-band images of the cluster\\
\textbf{Column 7}: Total $K_s$-band magnitude and error, given by the Auto (Kron) magnitude of SExtr (uncorrected for galactic extinction or $k$-correction for redshift)\\
\textbf{Column 8}: $B_J$ magnitude, corresponding approximately to $B_{25}$ isophotal magnitude, from the catalogue of Woudt \& Kraan-Korteweg (2001). The typical 1$\sigma$ uncertainty associated with these magnitudes is $0.^{\rm{m}}5$.\\
\textbf{Column 9}: Diameter of the object ($D$) in arcseconds, as estimated from SExtractor's {\ttfamily{ISOAREA}} parameter: $D \simeq 2 A_{\rm{iso}} = 2 \sqrt{\frac{\rm{ISOAREA} \times \rm{ELONGATION}}{\pi}}$ where $A_{\rm{iso}}$ is the isophotal semi-major axis, {\tt{ISOAREA}} is the isophotal area and {\ttfamily{ELONGATION}} is the ratio of the semi-major and semi-minor axes.\\
\textbf{Column 10}: Heliocentric velocity in km s$^{-1}$ from NED\footnote{The NASA/IPAC Extragalactic Database (NED) is operated by the Jet Propulsion Laboratory, California Institute of Technology, under contract with the National Aeronautics and Space Administration.}, with additional sources from 2dF spectroscopy (Woudt et al. 2009, in preparation).\\
\textbf{Column 11}: The field/s the object was identified in, corresponding to those shown in Fig. 1, followed by a photometric flag in brackets. The flag indicates the reliability of photometry in each image in which the object was detected. The number given is the sum of one or more of the following flags: ``0'' indicates reliable photometry, ``1'' indicates that neighbouring objects may have an influence on the photometry of the object, ``2'' indicates that the object was originally blended with another object, ``4'' indicates that at least one pixel is saturated, ``8'' indicates that the object was truncated due to proximity to the boundary of an image, ``16'' indicates that the object's aperture data is incomplete or corrupted, ``32'' indicates that the object's isophotal data is incomplete or corrupted. If an object is identified in more than one field, the magnitude given in Column 7 is the average of the data from all fields in which the photometric flag is zero.

\bsp
\scriptsize 
\onecolumn
\begin{longtable}[l]{lcccccccr@{  }lcrcl}
\label{cat.tbl}\\

\multicolumn{14}{l}{{\small{\bf{\tablename} \thetable{}.} The photometric catalogue of galaxies identified in the $K_s$-band}} \\[0.5ex]
\hline \\[-1.8ex]
$K_s$ ID & Optical ID & \multicolumn{2}{c}{RA (J2000)  Dec}& Gal $l$ & Gal $b$ & $A_{K_s}$ & Class & \multicolumn{2}{c} {$K_s$} & $B_J$ & D & Vel & Field/s (Flag)\\
& & [$^{h}$ $^{m}$ $^{s}$] & [$^{\circ}$ $^{\prime}$ $^{\prime\prime}$] & [$^{\circ}$]& [$^{\circ}$] & [mag] & & \multicolumn{2}{c} {[mag]} & [mag] & [$^{\prime\prime}$] & km s$^{-1}$ \\
(1) & (2) & (3a) & (3b) & (4a) &  (4b) & (5) & (6) & (7) & & (8) & (9) & (10) & (11)\\
\hline \\[-1.8ex]
\endfirsthead

\multicolumn{14}{l}{{\tablename} \thetable{} -- Continued} \\[0.5ex]
\hline \\[-1.8ex]
$K_s$ ID & Optical ID & \multicolumn{2}{c}{RA  (J2000)   Dec}& Gal $l$ & Gal $b$ & $A_{K_s}$ & Class & \multicolumn{2}{c} {$K_s$ [mag]} & $B_J$ & $D$ & Vel. & Field (Flag)\\
& & [$^{h}$ $^{m}$ $^{s}$] & [$^{\circ}$ $^{\prime}$ $^{\prime\prime}$] & [$^{\circ}$]& [$^{\circ}$] & [mag] & & \multicolumn{2}{c} {[mag]} & [mag] & [$^{\prime\prime}$] & km s$^{-1}$ \\
(1) & (2) & (3a) & (3b) & (4a) &  (4b) & (5) & (6) & (7) & & (8) & (9) & (10) & (11)\\
\hline \\[-1.8ex]
\endhead

\hline \\[-1.8ex]
\multicolumn{14}{l}{Continued on Next Page\ldots}\\
\endfoot

\hline
\endlastfoot

K001 & WKK6091 & 16 11 50.8 & -60 36 14 & 325.21 & -6.72 & 0.070 & G & 13.34 $\pm$ & 0.02 & 17.9 &   8 & 3235 & 29(0)
\\ 
K002 & WKK6092 & 16 11 51.4 & -60 37 55 & 325.20 & -6.74 & 0.070 & G & 10.74 $\pm$ & 0.00 & 14.7 &  26 & 4688 & 29(2)
\\ 
K003 & WKK6090 & 16 11 52.0 & -61 11 41 & 324.81 & -7.15 & 0.084 & G & 13.17 $\pm$ & 0.02 & 16.2 &   7 & ... & 34(2)
\\ 
K004 & ... & 16 11 52.1 & -60 37 34 & 325.20 & -6.74 & 0.070 & C & 15.04 $\pm$ & 0.04 & ... &   3 & ... & 29(0)
\\ 
K005 & ... & 16 11 52.6 & -61 09 09 & 324.84 & -7.12 & 0.086 & C & 15.39 $\pm$ & 0.05 & ... &   3 & ... & 34(0)
\\ 
K006 & ... & 16 11 54.2 & -60 37 18 & 325.21 & -6.74 & 0.070 & C & 16.33 $\pm$ & 0.07 & ... &   2 & ... & 29(0)
\\ 
K007 & ... & 16 11 55.6 & -60 39 39 & 325.18 & -6.77 & 0.074 & G & 15.19 $\pm$ & 0.05 & ... &   4 & ... & 29(0)
\\ 
K008 & WKK6100 & 16 11 55.9 & -60 43 33 & 325.14 & -6.82 & 0.079 & G & 12.36 $\pm$ & 0.01 & 15.7 &   9 & ... & 30(0)
\\ 
K009 & ... & 16 11 56.8 & -61 01 05 & 324.94 & -7.03 & 0.074 & C & 15.72 $\pm$ & 0.06 & ... &   2 & ... & 32(0)
\\ 
K010 & WKK6098 & 16 11 57.1 & -61 07 53 & 324.86 & -7.11 & 0.082 & G & 14.69 $\pm$ & 0.04 & 17.1 &   4 & 10482 & 33(0) 34(0)
\\
K011 & R & 16 11 57.6 & -61 10 07 & 324.83 & -7.14 & 0.086 & G & 14.57 $\pm$ & 0.03 & ... &   4 & ... & 34(0)
\\ 
K012 & WKK6101 & 16 11 58.6 & -60 51 11 & 325.05 & -6.92 & 0.081 & G & 12.51 $\pm$ & 0.01 & 16.7 &  15 & 4416 & 31(0)
\\ 
K013 & ... & 16 11 59.4 & -60 37 43 & 325.21 & -6.75 & 0.074 & G & 13.17 $\pm$ & 0.01 & ... &  11 & ... & 29(0)
\\ 
K014 & R & 16 12 00.4 & -60 51 26 & 325.05 & -6.92 & 0.081 & G & 13.71 $\pm$ & 0.02 & ... &   7 & ... & 31(0)
\\ 
K015 & ... & 16 12 01.0 & -61 01 15 & 324.94 & -7.04 & 0.074 & C & 15.68 $\pm$ & 0.06 & ... &   3 & ... & 33(0)
\\ 
K016 & R & 16 12 09.7 & -60 54 11 & 325.04 & -6.97 & 0.079 & G & 14.54 $\pm$ & 0.02 & ... &   5 & ... & 31(0) 32(0)
\\ 
K017 & R & 16 12 10.5 & -61 15 09 & 324.79 & -7.22 & 0.084 & G & 14.26 $\pm$ & 0.02 & ... &   6 & ... & 34(0)
\\ 
K018 & WKK6116 & 16 12 11.6 & -60 46 60 & 325.12 & -6.88 & 0.081 & G & 9.85 $\pm$ & 0.00 & 14.6 &  31 & 3803 & 30(0) 31(16)
\\ 
K019 & ... & 16 12 12.2 & -60 41 21 & 325.19 & -6.82 & 0.074 & C & 15.25 $\pm$ & 0.05 & ... &   3 & ... & 29(0) 30(0)
\\ 
K020 & R & 16 12 12.9 & -60 58 36 & 324.99 & -7.02 & 0.079 & G & 14.90 $\pm$ & 0.04 & ... &   4 & 16300 & 32(0)
\\ 
K021 & R & 16 12 13.3 & -61 05 52 & 324.91 & -7.11 & 0.082 & G & 14.48 $\pm$ & 0.03 & ... &   5 & ... & 33(0)
\\ 
K022 & ... & 16 12 13.6 & -60 34 12 & 325.27 & -6.73 & 0.076 & C & 15.84 $\pm$ & 0.06 & ... &   3 & ... & 29(0)
\\ 
K023 & R & 16 12 15.6 & -61 00 10 & 324.98 & -7.05 & 0.079 & G & 14.50 $\pm$ & 0.03 & ... &   8 & ... & 32(2)
\\ 
K024 & ... & 16 12 16.0 & -61 14 38 & 324.81 & -7.22 & 0.084 & C & 16.26 $\pm$ & 0.06 & ... &   3 & ... & 34(0)
\\ 
K025 & WKK6120 & 16 12 18.3 & -61 02 29 & 324.95 & -7.08 & 0.079 & G & 12.15 $\pm$ & 0.01 & 16.6 &  13 & 5508 & 33(0)
\\ 
K026 & ... & 16 12 18.5 & -60 44 37 & 325.16 & -6.86 & 0.076 & C & 15.85 $\pm$ & 0.04 & ... &   3 & ... & 30(0)
\\ 
K027 & WKK6123 & 16 12 20.2 & -61 06 38 & 324.91 & -7.13 & 0.082 & G & 11.66 $\pm$ & 0.00 & 14.7 &  15 & 10611 & 33(0)
\\ 
K028 & ... & 16 12 21.0 & -61 14 35 & 324.82 & -7.23 & 0.084 & C & 15.66 $\pm$ & 0.05 & ... &   2 & ... & 34(0)
\\ 
K029 & WKK6126 & 16 12 21.2 & -60 50 15 & 325.10 & -6.94 & 0.081 & G & 12.18 $\pm$ & 0.01 & 16.7 &  18 & 3770 & 31(0)
\\ 
K030 & WKK6125 & 16 12 22.3 & -61 04 48 & 324.93 & -7.11 & 0.082 & G & 12.80 $\pm$ & 0.01 & 16.9 &  12 & 3937 & 33(0)
\\ 
K031 & WKK6131 & 16 12 24.2 & -60 58 11 & 325.01 & -7.04 & 0.079 & G & 11.61 $\pm$ & 0.00 & 16.3 &  18 & 4784 & 32(0)
\\ 
K032 & R & 16 12 26.1 & -60 55 44 & 325.04 & -7.01 & 0.079 & G & 15.39 $\pm$ & 0.04 & ... &   3 & ... & 32(0)
\\ 
K033 & WKK6135 & 16 12 26.6 & -61 03 52 & 324.95 & -7.11 & 0.082 & G & 14.39 $\pm$ & 0.03 & 17.2 &   5 & 4426 & 33(0)
\\ 
K034 & ... & 16 12 30.3 & -60 42 43 & 325.20 & -6.86 & 0.076 & C & 16.05 $\pm$ & 0.06 & ... &   3 & ... & 30(0)
\\ 
K035 & ... & 16 12 31.2 & -61 00 35 & 324.99 & -7.07 & 0.079 & C & 15.97 $\pm$ & 0.06 & ... &   2 & ... & 32(0)
\\ 
K036 & ... & 16 12 32.5 & -60 52 31 & 325.09 & -6.98 & 0.077 & C & 15.55 $\pm$ & 0.05 & ... &   5 & ... & 31(0)
\\ 
K037 & ... & 16 12 33.3 & -60 34 43 & 325.30 & -6.76 & 0.075 & C & 16.02 $\pm$ & 0.06 & ... &   3 & ... & 29(0)
\\ 
K038 & ... & 16 12 34.4 & -60 48 03 & 325.14 & -6.93 & 0.081 & C & 15.41 $\pm$ & 0.05 & ... &   3 & ... & 31(0)
\\ 
K039 & ... & 16 12 36.7 & -60 43 44 & 325.20 & -6.88 & 0.076 & C & 16.48 $\pm$ & 0.08 & ... &   2 & ... & 30(0)
\\ 
K040 & ... & 16 12 37.7 & -61 07 38 & 324.92 & -7.17 & 0.084 & C & 15.55 $\pm$ & 0.05 & ... &   3 & ... & 34(0)
\\ 
K041 & ... & 16 12 39.4 & -61 13 51 & 324.85 & -7.25 & 0.084 & C & 15.77 $\pm$ & 0.06 & ... &   2 & ... & 34(0)
\\ 
K042 & R & 16 12 40.2 & -60 58 52 & 325.03 & -7.07 & 0.079 & G & 14.55 $\pm$ & 0.03 & ... &   5 & ... & 15(0)
\\ 
K043 & ... & 16 12 40.2 & -60 41 46 & 325.22 & -6.86 & 0.076 & C & 16.49 $\pm$ & 0.07 & ... &   2 & ... & 12(0)
\\ 
K044 & R & 16 12 40.2 & -60 58 52 & 325.03 & -7.07 & 0.079 & G & 14.43 $\pm$ & 0.02 & ... &   5 & ... & 32(0)
\\ 
K045 & ... & 16 12 40.3 & -61 03 59 & 324.97 & -7.13 & 0.079 & G & 14.95 $\pm$ & 0.03 & ... &   4 & ... & 15(0)
\\ 
K046 & ... & 16 12 41.6 & -60 50 06 & 325.13 & -6.96 & 0.077 & C & 15.01 $\pm$ & 0.04 & ... &   3 & ... & 31(0)
\\ 
K047 & ... & 16 12 42.2 & -60 49 41 & 325.14 & -6.96 & 0.077 & C & 16.53 $\pm$ & 0.08 & ... &   2 & ... & 31(0)
\\ 
K048 & ... & 16 12 42.6 & -61 11 36 & 324.88 & -7.22 & 0.084 & C & 16.28 $\pm$ & 0.07 & ... &   3 & ... & 16(0)
\\ 
K049 & ... & 16 12 42.7 & -60 38 51 & 325.26 & -6.83 & 0.075 & G & 15.89 $\pm$ & 0.04 & ... &   3 & ... & 12(0) 29(0)
\\ 
K050 & ... & 16 12 44.9 & -61 11 39 & 324.89 & -7.23 & 0.084 & C & 16.63 $\pm$ & 0.08 & ... &   2 & ... & 16(0)
\\ 
K051 & ... & 16 12 45.4 & -61 04 19 & 324.97 & -7.14 & 0.079 & G & 14.87 $\pm$ & 0.03 & ... &   4 & ... & 16(0) 33(0)
\\ 
K052 & ... & 16 12 45.6 & -61 11 08 & 324.89 & -7.22 & 0.084 & C & 15.96 $\pm$ & 0.05 & ... &   3 & ... & 16(0) 34(0)
\\ 
K053 & ... & 16 12 46.3 & -60 58 35 & 325.04 & -7.07 & 0.079 & C & 16.26 $\pm$ & 0.08 & ... &   2 & ... & 15(0)
\\ 
K054 & R & 16 12 47.7 & -60 39 34 & 325.26 & -6.84 & 0.075 & G & 13.73 $\pm$ & 0.01 & ... &   6 & ... & 12(0) 29(0)
\\ 
K055 & ... & 16 12 48.0 & -61 08 43 & 324.92 & -7.20 & 0.084 & C & 15.44 $\pm$ & 0.04 & ... &   4 & ... & 34(0)
\\ 
K056 & ... & 16 12 48.2 & -61 10 15 & 324.91 & -7.21 & 0.084 & C & 15.89 $\pm$ & 0.06 & ... &   4 & ... & 16(0)
\\ 
K057 & ... & 16 12 48.9 & -60 34 33 & 325.32 & -6.78 & 0.075 & C & 16.77 $\pm$ & 0.08 & ... &   2 & ... & 28(0)
\\ 
K058 & WKK6148 & 16 12 50.8 & -60 38 36 & 325.28 & -6.84 & 0.075 & G & 11.71 $\pm$ & 0.00 & 16.5 &  17 & 4272 & 12(0) 29(0)
\\ 
K059 & ... & 16 12 51.0 & -61 08 59 & 324.93 & -7.20 & 0.084 & C & 16.14 $\pm$ & 0.07 & ... &   2 & ... & 34(0)
\\ 
K060 & ... & 16 12 51.3 & -60 36 51 & 325.30 & -6.82 & 0.075 & G & 15.37 $\pm$ & 0.04 & ... &   4 & ... & 28(0) 29(0)
\\ 
K061 & ... & 16 12 51.4 & -61 11 49 & 324.89 & -7.24 & 0.084 & C & 16.19 $\pm$ & 0.07 & ... &   2 & ... & 34(0)
\\ 
K062 & ... & 16 12 51.7 & -61 11 57 & 324.89 & -7.24 & 0.084 & C & 15.62 $\pm$ & 0.05 & ... &   3 & ... & 16(0)
\\ 
K063 & R & 16 12 51.9 & -60 35 39 & 325.31 & -6.80 & 0.075 & G & 14.60 $\pm$ & 0.03 & ... &   4 & ... & 28(0) 29(0)
\\ 
K064 & ... & 16 12 52.8 & -60 35 24 & 325.32 & -6.80 & 0.075 & C & 15.85 $\pm$ & 0.05 & ... &   3 & ... & 28(0)
\\ 
K065 & ... & 16 12 54.4 & -60 38 38 & 325.28 & -6.84 & 0.075 & G & 14.21 $\pm$ & 0.01 & ... &   5 & ... & 12(0) 29(0)
\\ 
K066 & ... & 16 12 54.6 & -61 13 53 & 324.87 & -7.27 & 0.081 & C & 14.98 $\pm$ & 0.03 & ... &   5 & ... & 35(0)
\\ 
K067 & ... & 16 12 54.7 & -61 14 39 & 324.86 & -7.28 & 0.081 & C & 15.71 $\pm$ & 0.06 & ... &   3 & ... & 35(0)
\\ 
K068 & ... & 16 12 56.0 & -61 16 51 & 324.84 & -7.30 & 0.074 & C & 16.09 $\pm$ & 0.06 & ... &   3 & ... & 35(0)
\\ 
K069 & WKK6152 & 16 12 56.7 & -61 01 04 & 325.03 & -7.12 & 0.079 & G & 10.70 $\pm$ & 0.00 & 15.6 &  32 & 4266 & 15(3) 32(27) 33(19)
\\ 
K070 & ... & 16 12 58.1 & -60 36 57 & 325.31 & -6.83 & 0.075 & G & 14.27 $\pm$ & 0.01 & ... &   6 & ... & 12(0) 28(0)
\\ 
K071 & ... & 16 12 58.6 & -60 50 59 & 325.15 & -7.00 & 0.077 & C & 15.91 $\pm$ & 0.06 & ... &   3 & ... & 14(0)
\\ 
K072 & ... & 16 13 01.7 & -60 31 52 & 325.37 & -6.77 & 0.076 & C & 15.32 $\pm$ & 0.03 & ... &   3 & ... & 28(0)
\\ 
K073 & WKK6160 & 16 13 04.8 & -60 51 55 & 325.14 & -7.02 & 0.077 & G & 12.94 $\pm$ & 0.01 & 17.7 &  10 & 6508 & 14(0)
\\ 
K074 & ... & 16 13 05.9 & -61 14 00 & 324.89 & -7.28 & 0.081 & G & 15.22 $\pm$ & 0.04 & ... &   3 & ... & 35(0)
\\ 
K075 & ... & 16 13 06.4 & -61 07 57 & 324.96 & -7.21 & 0.081 & C & 15.41 $\pm$ & 0.05 & ... &   4 & ... & 16(0)
\\
K076 & R & 16 13 07.4 & -60 40 45 & 325.28 & -6.89 & 0.077 & G & 15.23 $\pm$ & 0.03 & ... &   4 & ... & 12(0)
\\ 
K077 & ... & 16 13 08.0 & -61 18 15 & 324.84 & -7.34 & 0.074 & G & 15.58 $\pm$ & 0.06 & ... &   3 & ... & 35(0)
\\ 
K078 & ... & 16 13 08.1 & -61 18 37 & 324.84 & -7.34 & 0.074 & G & 15.27 $\pm$ & 0.05 & ... &   3 & ... & 35(0)
\\ 
K079 & ... & 16 13 08.5 & -61 00 32 & 325.05 & -7.13 & 0.079 & C & 15.80 $\pm$ & 0.06 & ... &   2 & ... & 15(0)
\\ 
K080 & ... & 16 13 08.7 & -60 46 46 & 325.21 & -6.96 & 0.073 & C & 16.44 $\pm$ & 0.07 & ... &   2 & ... & 13(0) 13(0)
\\ 
K081 & ... & 16 13 09.4 & -61 13 28 & 324.90 & -7.28 & 0.081 & C & 15.01 $\pm$ & 0.03 & ... &   4 & ... & 35(0)
\\ 
K082 & R & 16 13 10.0 & -60 36 56 & 325.32 & -6.84 & 0.075 & G & 15.09 $\pm$ & 0.02 & ... &   4 & ... & 28(0)
\\ 
K083 & WKK6166 & 16 13 10.1 & -60 45 04 & 325.23 & -6.94 & 0.076 & G & 13.25 $\pm$ & 0.01 & 17.1 &   9 & 5073 & 12(24) 13(2)
\\ 
K084 & ... & 16 13 13.4 & -61 05 23 & 325.00 & -7.19 & 0.081 & C & 16.25 $\pm$ & 0.04 & ... &   3 & ... & 16(0)
\\ 
K085 & ... & 16 13 15.1 & -61 01 01 & 325.05 & -7.14 & 0.076 & C & 15.67 $\pm$ & 0.05 & ... &   3 & ... & 15(0)
\\ 
K086 & R & 16 13 16.8 & -60 54 36 & 325.13 & -7.07 & 0.076 & G & 15.96 $\pm$ & 0.07 & ... &   3 & ... & 14(0)
\\ 
K087 & ... & 16 13 21.0 & -60 40 03 & 325.31 & -6.90 & 0.077 & C & 16.35 $\pm$ & 0.07 & ... &   2 & ... & 12(0)
\\ 
K088 & R & 16 13 21.2 & -60 50 14 & 325.19 & -7.02 & 0.073 & G & 15.10 $\pm$ & 0.04 & ... &   3 & ... & 13(0)
\\ 
K089 & ... & 16 13 24.9 & -61 13 33 & 324.92 & -7.30 & 0.081 & C & 14.45 $\pm$ & 0.01 & ... &   5 & ... & 35(0)
\\ 
K090 & ... & 16 13 25.9 & -60 43 45 & 325.27 & -6.95 & 0.077 & G & 16.14 $\pm$ & 0.06 & ... &   2 & ... & 12(0)
\\ 
K091 & ... & 16 13 26.3 & -61 08 47 & 324.98 & -7.25 & 0.081 & C & 15.59 $\pm$ & 0.05 & ... &   3 & ... & 16(2)
\\ 
K092 & R & 16 13 26.4 & -60 44 21 & 325.26 & -6.96 & 0.080 & G & 14.07 $\pm$ & 0.03 & ... &   7 & ... & 12(0) 13(0)
\\ 
K093 & R & 16 13 27.2 & -60 48 55 & 325.21 & -7.01 & 0.073 & G & 15.76 $\pm$ & 0.05 & ... &   3 & ... & 13(0)
\\ 
K094 & WKK6176 & 16 13 27.2 & -60 45 50 & 325.25 & -6.97 & 0.073 & G & 11.62 $\pm$ & 0.01 & 14.6 &  19 & 4680 & 13(0)
\\ 
K095 & ... & 16 13 30.7 & -60 35 24 & 325.37 & -6.85 & 0.078 & C & 16.02 $\pm$ & 0.06 & ... &   2 & ... & 28(0)
\\ 
K096 & WKK6180 & 16 13 32.2 & -61 00 22 & 325.09 & -7.16 & 0.080 & G & 10.55 $\pm$ & 0.00 & 15.2 &  24 & 4602 & 5(18) 15(2)
\\ 
K097 & R & 16 13 32.3 & -60 51 41 & 325.19 & -7.05 & 0.073 & G & 14.17 $\pm$ & 0.03 & ... &   4 & ... & 14(2)
\\ 
K098 & WKK6183 & 16 13 32.9 & -60 49 23 & 325.21 & -7.03 & 0.073 & G & 10.44 $\pm$ & 0.00 & 15.6 &  25 & 5909 & 3(18) 13(0)
\\ 
K099 & R & 16 13 34.5 & -60 51 05 & 325.20 & -7.05 & 0.073 & G & 12.94 $\pm$ & 0.01 & ... &   8 & 4583 & 3(3) 4(3) 13(2) 14(2)
\\ 
K100 & WKK6186 & 16 13 35.7 & -60 51 54 & 325.19 & -7.06 & 0.073 & G & 9.91 $\pm$ & 0.00 & 14.8 &  50 & 5743 & 3(27) 4(18) 14(2)
\\ 
K101 & R & 16 13 36.5 & -60 52 49 & 325.18 & -7.07 & 0.073 & G & 14.83 $\pm$ & 0.04 & ... &   4 & 4117 & 14(0)
\\ 
K102 & ... & 16 13 36.9 & -60 30 45 & 325.44 & -6.81 & 0.076 & C & 15.57 $\pm$ & 0.05 & ... &   3 & ... & 28(0)
\\ 
K103 & R & 16 13 37.1 & -61 04 45 & 325.04 & -7.22 & 0.081 & G & 15.58 $\pm$ & 0.05 & ... &   3 & ... & 5(0) 15(0) 16(0) 17(0)
\\ 
K104 & ... & 16 13 38.3 & -60 35 49 & 325.38 & -6.87 & 0.078 & C & 16.77 $\pm$ & 0.08 & ... &   2 & ... & 28(0)
\\ 
K105 & R & 16 13 39.0 & -60 57 20 & 325.13 & -7.13 & 0.076 & G & 15.27 $\pm$ & 0.04 & ... &   3 & ... & 5(0) 14(0) 15(0)
\\ 
K106 & WKK6190 & 16 13 39.4 & -60 59 56 & 325.10 & -7.16 & 0.080 & G & 11.08 $\pm$ & 0.00 & 16.0 &  19 & 4309 & 5(0) 15(0)
\\ 
K107 & ... & 16 13 40.4 & -60 38 26 & 325.35 & -6.90 & 0.078 & C & 15.50 $\pm$ & 0.04 & ... &   3 & ... & 11(0) 12(0)
\\ 
K108 & ... & 16 13 40.4 & -61 12 12 & 324.96 & -7.31 & 0.083 & C & 15.93 $\pm$ & 0.07 & ... &   2 & ... & 36(0)
\\ 
K109 & ... & 16 13 41.3 & -61 17 34 & 324.90 & -7.38 & 0.072 & C & 15.58 $\pm$ & 0.06 & ... &   3 & ... & 36(0)
\\ 
K110 & ... & 16 13 41.6 & -60 44 30 & 325.28 & -6.98 & 0.080 & C & 15.70 $\pm$ & 0.06 & ... &   2 & ... & 3(0)
\\ 
K111 & R & 16 13 41.8 & -61 14 19 & 324.94 & -7.34 & 0.081 & G & 14.41 $\pm$ & 0.02 & ... &   5 & ... & 35(0)
\\ 
K112 & ... & 16 13 42.2 & -60 37 01 & 325.37 & -6.89 & 0.078 & G & 15.37 $\pm$ & 0.04 & ... &   3 & ... & 11(0)
\\ 
K113 & R & 16 13 44.4 & -60 55 01 & 325.17 & -7.11 & 0.073 & G & 13.90 $\pm$ & 0.02 & ... &   6 & 5705 & 4(0) 14(16)
\\ 
K114 & ... & 16 13 45.6 & -61 00 01 & 325.11 & -7.17 & 0.080 & C & 12.45 $\pm$ & 0.01 & ... &  12 & ... & 5(3)
\\ 
K115 & WKK6193 & 16 13 46.0 & -61 00 29 & 325.10 & -7.18 & 0.080 & G & 10.06 $\pm$ & 0.00 & 14.9 &  44 & 4235 & 5(3)
\\ 
K116 & WKK6196 & 16 13 47.6 & -60 53 04 & 325.19 & -7.09 & 0.073 & G & 11.71 $\pm$ & 0.01 & 16.2 &  14 & 4818 & 4(3)
\\ 
K117 & R & 16 13 50.1 & -61 10 09 & 325.00 & -7.30 & 0.083 & G & 13.97 $\pm$ & 0.03 & ... &   9 & ... & 17(3)
\\ 
K118 & WKK6199 & 16 13 50.2 & -60 37 22 & 325.38 & -6.91 & 0.078 & G & 11.60 $\pm$ & 0.00 & 16.2 &  24 & ... & 11(2) 27(0)
\\ 
K119 & R & 16 13 50.6 & -60 57 45 & 325.14 & -7.15 & 0.080 & G & 16.16 $\pm$ & 0.07 & ... &   2 & ... & 4(0) 5(0)
\\ 
K120 & ... & 16 13 53.1 & -60 51 19 & 325.22 & -7.08 & 0.073 & C & 16.26 $\pm$ & 0.07 & ... &   3 & ... & 3(0)
\\ 
K121 & WKK6202 & 16 13 53.5 & -60 47 49 & 325.26 & -7.04 & 0.080 & G & 11.32 $\pm$ & 0.01 & 16.1 &  22 & 3521 & 3(0)
\\ 
K122 & WKK62** NC0656 & 16 13 53.7 & -60 52 02 & 325.21 & -7.09 & 0.073 & G & 10.67 $\pm$ & 0.00 & 15.4 &  36 & 3007 & 4(2)
\\ 
K123 & ... & 16 13 55.3 & -61 03 27 & 325.08 & -7.23 & 0.080 & C & 15.30 $\pm$ & 0.05 & ... &   3 & ... & 5(0)
\\ 
K124 & WKK6204 & 16 13 56.2 & -61 00 41 & 325.12 & -7.19 & 0.080 & G & 9.49 $\pm$ & 0.00 & 14.9 &  40 & 4584 & 5(3)
\\ 
K125 & WKK6207 & 16 13 57.5 & -60 56 24 & 325.17 & -7.14 & 0.073 & G & 10.39 $\pm$ & 0.00 & 15.3 &  34 & 3468 & 4(2)
\\ 
K126 & WKK6209 & 16 13 58.6 & -60 54 27 & 325.19 & -7.12 & 0.073 & G & 13.30 $\pm$ & 0.02 & 18.1 &   7 & ... & 4(2)
\\ 
K127 & ... & 16 13 59.0 & -60 46 23 & 325.29 & -7.03 & 0.080 & G & 13.60 $\pm$ & 0.02 & ... &   9 & ... & 3(0)
\\ 
K128 & ... & 16 13 59.8 & -60 47 04 & 325.28 & -7.04 & 0.080 & C & 16.37 $\pm$ & 0.07 & ... &   2 & ... & 3(0)
\\ 
K129 & WKK6211 & 16 14 00.0 & -60 58 39 & 325.15 & -7.17 & 0.080 & G & 12.98 $\pm$ & 0.01 & 17.4 &  11 & 4520 & 4(16) 5(0)
\\ 
K130 & WKK6212 & 16 14 01.0 & -60 59 28 & 325.14 & -7.19 & 0.080 & G & 10.22 $\pm$ & 0.00 & 15.6 &  28 & 3323 & 5(2)
\\ 
K131 & R & 16 14 01.8 & -60 59 40 & 325.14 & -7.19 & 0.080 & G & 13.14 $\pm$ & 0.02 & ... &  10 & ... & 5(3)
\\ 
K132 & WKK6216 & 16 14 02.1 & -60 54 05 & 325.20 & -7.12 & 0.073 & G & 12.66 $\pm$ & 0.01 & 17.1 &  12 & 5056 & 4(0)
\\ 
K133 & ... & 16 14 02.6 & -61 02 50 & 325.10 & -7.23 & 0.080 & C & 15.98 $\pm$ & 0.06 & ... &   3 & ... & 5(0)
\\ 
K134 & WKK6214 & 16 14 02.7 & -61 13 04 & 324.98 & -7.35 & 0.076 & G & 13.67 $\pm$ & 0.02 & 17.1 &   8 & 5054 & 36(0)
\\ 
K135 & R & 16 14 02.8 & -61 00 54 & 325.12 & -7.21 & 0.080 & G & 13.66 $\pm$ & 0.02 & ... &   7 & 4501 & 5(0)
\\ 
K136 & ... & 16 14 03.2 & -60 58 28 & 325.15 & -7.18 & 0.074 & C & 16.39 $\pm$ & 0.08 & ... &   2 & ... & 5(0)
\\ 
K137 & ... & 16 14 03.3 & -61 10 55 & 325.01 & -7.33 & 0.083 & C & 16.18 $\pm$ & 0.07 & ... &   2 & ... & 17(0)
\\ 
K138 & R & 16 14 04.4 & -60 46 55 & 325.29 & -7.04 & 0.080 & G & 14.43 $\pm$ & 0.03 & ... &   4 & 3564 & 3(0)
\\ 
K139 & ... & 16 14 05.4 & -60 34 25 & 325.44 & -6.89 & 0.078 & C & 15.85 $\pm$ & 0.04 & ... &   2 & ... & 27(0)
\\ 
K140 & ... & 16 14 05.7 & -61 04 11 & 325.09 & -7.25 & 0.080 & C & 15.72 $\pm$ & 0.05 & ... &   3 & ... & 5(2)
\\ 
K141 & ... & 16 14 05.8 & -60 50 45 & 325.25 & -7.09 & 0.073 & C & 14.87 $\pm$ & 0.02 & ... &   5 & ... & 3(0)
\\ 
K142 & WKK6228 & 16 14 06.1 & -60 41 54 & 325.35 & -6.98 & 0.083 & G & 10.93 $\pm$ & 0.00 & 16.0 &  28 & 5671 & 11(3)
\\ 
K143 & ... & 16 14 07.0 & -61 07 19 & 325.06 & -7.29 & 0.083 & G & 14.31 $\pm$ & 0.02 & ... &   6 & ... & 17(0)
\\ 
K144 & R & 16 14 07.7 & -60 49 08 & 325.27 & -7.07 & 0.076 & G & 14.66 $\pm$ & 0.04 & ... &   4 & 4647 & 3(2)
\\ 
K145 & R & 16 14 08.6 & -61 03 20 & 325.10 & -7.24 & 0.080 & G & 13.90 $\pm$ & 0.02 & ... &   6 & 5185 & 5(0)
\\ 
K146 & ... & 16 14 08.7 & -61 11 10 & 325.01 & -7.34 & 0.083 & C & 15.22 $\pm$ & 0.03 & ... &   3 & ... & 17(3)
\\ 
K147 & ... & 16 14 09.2 & -61 09 21 & 325.03 & -7.32 & 0.083 & C & 14.50 $\pm$ & 0.01 & ... &   5 & ... & 17(0)
\\ 
K148 & ... & 16 14 09.6 & -60 45 54 & 325.31 & -7.04 & 0.080 & C & 15.84 $\pm$ & 0.06 & ... &   3 & ... & 3(0)
\\ 
K149 & ... & 16 14 09.8 & -61 13 57 & 324.98 & -7.37 & 0.076 & C & 15.34 $\pm$ & 0.04 & ... &   4 & ... & 36(0)
\\ 
K150 & WKK6229 & 16 14 10.4 & -60 51 01 & 325.25 & -7.10 & 0.076 & G & 11.58 $\pm$ & 0.00 & 16.6 &  13 & 5376 & 3(0) 4(0)
\\
K151 & ... & 16 14 10.6 & -60 58 18 & 325.17 & -7.19 & 0.074 & C & 15.40 $\pm$ & 0.05 & ... &   3 & ... & 5(0)
\\ 
K152 & WKK6230 & 16 14 10.7 & -60 44 32 & 325.33 & -7.02 & 0.080 & G & 11.44 $\pm$ & 0.00 & 16.6 &  22 & 3025 & 11(2) 3(2)
\\ 
K153 & ... & 16 14 13.5 & -61 05 33 & 325.09 & -7.28 & 0.083 & G & 12.84 $\pm$ & 0.01 & ... &   9 & ... & 17(0)
\\ 
K154 & R & 16 14 15.3 & -61 00 28 & 325.15 & -7.22 & 0.080 & G & 15.05 $\pm$ & 0.04 & ... &   3 & ... & 5(0)
\\ 
K155 & ... & 16 14 15.8 & -60 47 34 & 325.30 & -7.06 & 0.076 & C & 15.58 $\pm$ & 0.06 & ... &   3 & ... & 3(0)
\\ 
K156 & WKK6233 & 16 14 18.0 & -60 53 26 & 325.23 & -7.14 & 0.073 & G & 11.83 $\pm$ & 0.00 & 17.1 &  13 & 5055 & 4(0)
\\ 
K157 & R & 16 14 19.2 & -61 05 09 & 325.10 & -7.28 & 0.083 & G & 14.78 $\pm$ & 0.03 & ... &   6 & ... & 5(16) 17(0)
\\ 
K158 & R & 16 14 19.8 & -61 04 58 & 325.10 & -7.28 & 0.083 & G & 14.06 $\pm$ & 0.02 & ... &   7 & 29515 & 5(0) 17(0)
\\ 
K159 & ... & 16 14 20.5 & -61 17 12 & 324.96 & -7.43 & 0.076 & C & 15.63 $\pm$ & 0.06 & ... &   3 & ... & 36(0)
\\ 
K160 & WKK6235 & 16 14 22.6 & -61 08 38 & 325.06 & -7.33 & 0.083 & G & 11.04 $\pm$ & 0.00 & 15.6 &  21 & 4053 & 17(0)
\\ 
K161 & ... & 16 14 23.0 & -61 13 59 & 325.00 & -7.39 & 0.076 & C & 16.02 $\pm$ & 0.07 & ... &   2 & ... & 36(0)
\\ 
K162 & ... & 16 14 24.1 & -60 44 38 & 325.35 & -7.04 & 0.080 & C & 15.82 $\pm$ & 0.04 & ... &   3 & ... & 11(0)
\\ 
K163 & WKK6239 & 16 14 25.4 & -61 03 40 & 325.12 & -7.27 & 0.083 & G & 10.48 $\pm$ & 0.00 & 15.2 &  34 & 4235 & 5(2)
\\ 
K164 & ... & 16 14 29.4 & -60 43 20 & 325.37 & -7.03 & 0.083 & C & 15.97 $\pm$ & 0.04 & ... &   3 & ... & 11(0)
\\ 
K165 & ... & 16 14 29.4 & -61 04 52 & 325.12 & -7.29 & 0.083 & G & 15.08 $\pm$ & 0.04 & ... &   4 & ... & 5(0) 17(0) 18(2)
\\ 
K166 & ... & 16 14 30.0 & -61 01 28 & 325.16 & -7.25 & 0.077 & C & 15.78 $\pm$ & 0.04 & ... &   3 & ... & 5(0)
\\ 
K167 & WKK6242 & 16 14 30.5 & -60 53 46 & 325.25 & -7.16 & 0.074 & G & 10.71 $\pm$ & 0.00 & 16.4 &  21 & 5197 & 1(0) 4(2)
\\ 
K168 & ... & 16 14 30.5 & -60 41 11 & 325.39 & -7.01 & 0.083 & C & 16.88 $\pm$ & 0.08 & ... &   2 & ... & 11(0)
\\ 
K169 & ... & 16 14 30.8 & -61 09 56 & 325.06 & -7.35 & 0.079 & G & 15.91 $\pm$ & 0.06 & ... &   2 & ... & 17(0)
\\ 
K170 & ... & 16 14 31.8 & -61 11 09 & 325.05 & -7.37 & 0.076 & C & 15.95 $\pm$ & 0.05 & ... &   3 & ... & 37(0)
\\ 
K171 & ... & 16 14 33.3 & -61 17 03 & 324.98 & -7.44 & 0.076 & G & 15.42 $\pm$ & 0.05 & ... &   3 & ... & 36(0) 37(3)
\\ 
K172 & ... & 16 14 33.6 & -61 01 48 & 325.16 & -7.26 & 0.077 & C & 15.97 $\pm$ & 0.05 & ... &   3 & ... & 5(0)
\\ 
K173 & R & 16 14 33.7 & -60 54 16 & 325.25 & -7.17 & 0.074 & G & 14.19 $\pm$ & 0.02 & ... &   5 & 29400 & 1(0)
\\ 
K174 & ... & 16 14 33.8 & -61 03 56 & 325.13 & -7.29 & 0.083 & G & 15.44 $\pm$ & 0.04 & ... &   3 & ... & 5(0)
\\ 
K175 & R & 16 14 36.3 & -60 45 04 & 325.36 & -7.06 & 0.087 & G & 15.11 $\pm$ & 0.04 & ... &   4 & ... & 2(0) 10(16)
\\ 
K176 & R & 16 14 37.5 & -61 05 04 & 325.13 & -7.30 & 0.083 & G & 14.64 $\pm$ & 0.03 & ... &   6 & ... & 6(3) 5(3) 17(3) 18(3)
\\ 
K177 & R & 16 14 38.6 & -61 11 01 & 325.06 & -7.38 & 0.079 & G & 16.09 $\pm$ & 0.06 & ... &   2 & ... & 18(0)
\\ 
K178 & ... & 16 14 40.0 & -60 49 46 & 325.31 & -7.12 & 0.076 & C & 15.79 $\pm$ & 0.06 & ... &   3 & ... & 2(0)
\\ 
K179 & R & 16 14 40.7 & -61 03 30 & 325.15 & -7.29 & 0.083 & G & 15.82 $\pm$ & 0.04 & ... &   3 & ... & 5(16)
\\ 
K180 & WKK6248 & 16 14 41.1 & -60 35 39 & 325.48 & -6.96 & 0.089 & G & 14.47 $\pm$ & 0.03 & 17.7 &   7 & ... & 26(0) 27(16)
\\ 
K181 & ... & 16 14 43.1 & -61 16 40 & 325.00 & -7.45 & 0.074 & G & 14.65 $\pm$ & 0.04 & ... &   4 & ... & 37(2)
\\ 
K182 & ... & 16 14 43.2 & -61 04 03 & 325.15 & -7.30 & 0.083 & C & 15.64 $\pm$ & 0.04 & ... &   3 & ... & 6(0)
\\ 
K183 & ... & 16 14 44.0 & -61 09 50 & 325.08 & -7.37 & 0.079 & C & 16.50 $\pm$ & 0.08 & ... &   2 & ... & 18(0)
\\ 
K184 & ... & 16 14 44.1 & -60 50 03 & 325.31 & -7.13 & 0.076 & C & 16.14 $\pm$ & 0.07 & ... &   2 & ... & 2(0)
\\ 
K185 & WKK6251 & 16 14 45.1 & -60 55 36 & 325.25 & -7.20 & 0.074 & G & 11.47 $\pm$ & 0.01 & 15.9 &  24 & 5689 & 1(0)
\\ 
K186 & WKK6250 & 16 14 45.2 & -61 01 50 & 325.18 & -7.28 & 0.077 & G & 10.59 $\pm$ & 0.00 & 15.7 &  28 & 6258 & 6(0)
\\ 
K187 & WKK6252 & 16 14 45.4 & -60 51 41 & 325.29 & -7.16 & 0.072 & G & 12.10 $\pm$ & 0.01 & 16.8 &  13 & 5863 & 1(0) 2(0)
\\ 
K188 & R & 16 14 46.5 & -60 53 59 & 325.27 & -7.18 & 0.072 & G & 16.00 $\pm$ & 0.08 & ... &   2 & ... & 1(0)
\\ 
K189 & WKK6254 & 16 14 47.6 & -61 11 43 & 325.06 & -7.40 & 0.079 & G & 11.51 $\pm$ & 0.00 & 15.4 &  27 & 4910 & 18(2) 37(0)
\\ 
K190 & ... & 16 14 48.8 & -60 54 20 & 325.27 & -7.19 & 0.072 & G & 15.68 $\pm$ & 0.06 & ... &   3 & ... & 1(0)
\\ 
K191 & ... & 16 14 49.5 & -60 37 26 & 325.47 & -6.99 & 0.089 & C & 16.11 $\pm$ & 0.04 & ... &   3 & ... & 10(0)
\\ 
K192 & ... & 16 14 51.1 & -60 57 42 & 325.23 & -7.24 & 0.074 & C & 15.14 $\pm$ & 0.04 & ... &   3 & ... & 1(0)
\\ 
K193 & ... & 16 14 52.2 & -60 52 49 & 325.29 & -7.18 & 0.072 & G & 15.48 $\pm$ & 0.06 & ... &   3 & ... & 1(0)
\\ 
K194 & ... & 16 14 55.0 & -60 59 50 & 325.21 & -7.27 & 0.077 & G & 13.57 $\pm$ & 0.02 & ... &   5 & ... & 6(3)
\\ 
K195 & ... & 16 14 56.9 & -60 42 26 & 325.42 & -7.06 & 0.087 & C & 15.90 $\pm$ & 0.04 & ... &   3 & ... & 10(0)
\\ 
K196 & ... & 16 14 58.0 & -60 41 21 & 325.43 & -7.05 & 0.083 & C & 16.15 $\pm$ & 0.07 & ... &   2 & ... & 10(0)
\\ 
K197 & WKK6263 & 16 14 58.9 & -60 46 04 & 325.38 & -7.11 & 0.087 & G & 13.95 $\pm$ & 0.03 & 17.2 &   4 & 3260 & 2(0)
\\ 
K198 & WKK6264 & 16 14 59.6 & -60 52 01 & 325.31 & -7.18 & 0.072 & G & 13.57 $\pm$ & 0.02 & 17.8 &   7 & 4940 & 1(2)
\\ 
K199 & WKK6265 & 16 14 60.0 & -60 47 23 & 325.37 & -7.12 & 0.087 & G & 13.42 $\pm$ & 0.02 & 16.4 &   9 & 5611 & 2(2)
\\ 
K200 & R & 16 15 00.2 & -61 01 05 & 325.21 & -7.29 & 0.077 & G & 13.61 $\pm$ & 0.01 & ... &   8 & ... & 6(0)
\\ 
K201 & ... & 16 15 00.4 & -60 37 53 & 325.48 & -7.01 & 0.089 & C & 16.26 $\pm$ & 0.07 & ... &   2 & ... & 10(0)
\\ 
K202 & ... & 16 15 00.5 & -60 44 25 & 325.40 & -7.09 & 0.087 & G & 16.18 $\pm$ & 0.07 & ... &   2 & ... & 2(0) 10(0)
\\ 
K203 & WKK6267 & 16 15 01.1 & -60 48 34 & 325.35 & -7.14 & 0.087 & G & 14.38 $\pm$ & 0.03 & 17.9 &   4 & ... & 2(0)
\\ 
K204 & ... & 16 15 03.4 & -61 03 58 & 325.18 & -7.33 & 0.077 & G & 16.01 $\pm$ & 0.07 & ... &   3 & ... & 6(0)
\\ 
K205 & WKK6269 & 16 15 03.8 & -60 54 26 & 325.29 & -7.21 & 0.072 & G & 8.56 $\pm$ & 0.00 & 13.5 &  59 & 5441 & 1(2)
\\ 
K206 & ... & 16 15 07.2 & -60 57 42 & 325.26 & -7.26 & 0.072 & C & 16.04 $\pm$ & 0.07 & ... &   3 & ... & 1(0)
\\ 
K207 & R & 16 15 09.4 & -60 59 29 & 325.24 & -7.28 & 0.077 & G & 15.84 $\pm$ & 0.06 & ... &   2 & ... & 6(0)
\\ 
K208 & WKK6275 & 16 15 09.9 & -60 53 28 & 325.31 & -7.21 & 0.072 & G & 10.65 $\pm$ & 0.00 & 15.3 &  29 & 6861 & 1(2)
\\ 
K209 & R & 16 15 10.5 & -60 52 31 & 325.32 & -7.20 & 0.072 & G & 13.81 $\pm$ & 0.02 & ... &   6 & 6423 & 1(0)
\\ 
K210 & WKK6277 & 16 15 11.4 & -61 12 04 & 325.09 & -7.44 & 0.079 & G & 14.38 $\pm$ & 0.03 & 18.2 &   6 & 4602 & 18(16) 37(0)
\\ 
K211 & WKK6278 & 16 15 11.8 & -60 31 52 & 325.56 & -6.96 & 0.088 & G & 13.54 $\pm$ & 0.02 & 17.2 &  10 & 4593 & 26(0)
\\ 
K212 & ... & 16 15 11.8 & -60 52 01 & 325.33 & -7.20 & 0.072 & C & 13.80 $\pm$ & 0.01 & ... &   6 & ... & 1(3)
\\ 
K213 & ... & 16 15 12.9 & -61 16 35 & 325.04 & -7.49 & 0.074 & C & 15.90 $\pm$ & 0.06 & ... &   3 & ... & 37(0)
\\ 
K214 & ... & 16 15 13.5 & -60 38 23 & 325.49 & -7.04 & 0.089 & C & 14.96 $\pm$ & 0.02 & ... &   4 & ... & 10(0)
\\ 
K215 & ... & 16 15 14.2 & -61 00 59 & 325.23 & -7.31 & 0.077 & C & 16.50 $\pm$ & 0.08 & ... &   2 & ... & 6(0)
\\ 
K216 & R & 16 15 14.7 & -60 37 59 & 325.50 & -7.03 & 0.089 & G & 14.22 $\pm$ & 0.01 & ... &   5 & ... & 10(0) 26(0)
\\ 
K217 & WKK6283 & 16 15 14.9 & -60 35 31 & 325.53 & -7.00 & 0.089 & G & 10.74 $\pm$ & 0.00 & 15.2 &  26 & 6225 & 26(2)
\\ 
K218 & WKK6282 & 16 15 15.5 & -60 56 16 & 325.28 & -7.25 & 0.072 & G & 11.19 $\pm$ & 0.00 & 15.9 &  16 & 4849 & 1(2)
\\ 
K219 & R & 16 15 15.8 & -60 37 51 & 325.50 & -7.03 & 0.089 & G & 15.19 $\pm$ & 0.04 & ... &   4 & ... & 10(0)
\\ 
K220 & ... & 16 15 16.0 & -60 57 05 & 325.28 & -7.26 & 0.072 & C & 15.97 $\pm$ & 0.07 & ... &   3 & ... & 1(0)
\\ 
K221 & WKK6281 & 16 15 16.0 & -61 07 58 & 325.15 & -7.39 & 0.079 & G & 10.87 $\pm$ & 0.00 & 15.5 &  37 & 4908 & 18(3)
\\ 
K222 & WKK6288 & 16 15 17.9 & -60 45 06 & 325.42 & -7.12 & 0.087 & G & 13.80 $\pm$ & 0.02 & 17.7 &   8 & ... & 2(0)
\\ 
K223 & ... & 16 15 18.0 & -60 39 28 & 325.49 & -7.06 & 0.098 & G & 14.84 $\pm$ & 0.02 & ... &   5 & ... & 10(0)
\\ 
K224 & ... & 16 15 19.3 & -60 58 17 & 325.27 & -7.28 & 0.072 & C & 15.85 $\pm$ & 0.05 & ... &   3 & ... & 1(3)
\\ 
K225 & ... & 16 15 19.3 & -60 45 14 & 325.42 & -7.13 & 0.087 & C & 15.64 $\pm$ & 0.06 & ... &   2 & ... & 2(0)
\\
K226 & WKK6289 & 16 15 19.9 & -61 03 25 & 325.21 & -7.34 & 0.077 & G & 12.71 $\pm$ & 0.01 & 17.6 &  10 & 4447 & 6(2)
\\ 
K227 & ... & 16 15 21.9 & -61 13 35 & 325.09 & -7.47 & 0.074 & C & 15.48 $\pm$ & 0.04 & ... &   5 & ... & 37(0)
\\ 
K228 & WKK6293 & 16 15 23.4 & -60 42 48 & 325.45 & -7.10 & 0.098 & G & 14.31 $\pm$ & 0.02 & 18.3 &   8 & ... & 10(0)
\\ 
K229 & WKK6291 & 16 15 23.5 & -60 52 41 & 325.34 & -7.22 & 0.072 & G & 14.05 $\pm$ & 0.03 & 17.8 &   5 & ... & 1(0) 1(0) 8(2)
\\ 
K230 & ... & 16 15 24.9 & -60 57 44 & 325.28 & -7.28 & 0.072 & C & 14.81 $\pm$ & 0.02 & ... &   4 & ... & 6(0)
\\ 
K231 & WKK6294 & 16 15 25.2 & -61 04 11 & 325.21 & -7.36 & 0.077 & G & 12.08 $\pm$ & 0.01 & 16.8 &  16 & 4547 & 6(0) 7(0) 18(0) 19(0)
\\ 
K232 & ... & 16 15 25.2 & -60 31 31 & 325.59 & -6.97 & 0.088 & C & 16.36 $\pm$ & 0.07 & ... &   2 & ... & 47(0)
\\ 
K233 & WKK6295 & 16 15 25.9 & -61 00 22 & 325.25 & -7.32 & 0.072 & G & 13.17 $\pm$ & 0.01 & 17.6 &   7 & 6101 & 6(2) 7(2)
\\ 
K234 & ... & 16 15 26.4 & -61 06 08 & 325.19 & -7.39 & 0.077 & C & 15.93 $\pm$ & 0.05 & ... &   3 & ... & 19(0)
\\ 
K235 & WKK6297 & 16 15 27.1 & -60 58 50 & 325.27 & -7.30 & 0.072 & G & 11.79 $\pm$ & 0.01 & 16.2 &  16 & 5023 & 6(0) 7(2)
\\ 
K236 & ... & 16 15 27.7 & -61 12 25 & 325.11 & -7.46 & 0.074 & C & 16.65 $\pm$ & 0.10 & ... &   2 & ... & 38(0)
\\ 
K237 & WKK6299 & 16 15 28.9 & -60 48 56 & 325.39 & -7.18 & 0.078 & G & 11.53 $\pm$ & 0.01 & 17.4 &  15 & 5211 & 2(0)
\\ 
K238 & WKK6300 & 16 15 29.3 & -60 51 03 & 325.37 & -7.21 & 0.078 & G & 12.98 $\pm$ & 0.01 & 17.4 &   7 & 4586 & 1(0) 2(0) 8(0) 9(0)
\\ 
K239 & ... & 16 15 30.1 & -60 45 42 & 325.43 & -7.15 & 0.087 & C & 14.70 $\pm$ & 0.03 & ... &   4 & ... & 9(0)
\\ 
K240 & ... & 16 15 30.5 & -60 42 35 & 325.47 & -7.11 & 0.098 & C & 15.87 $\pm$ & 0.06 & ... &   3 & ... & 10(0)
\\ 
K241 & ... & 16 15 30.9 & -60 45 12 & 325.44 & -7.14 & 0.087 & C & 16.05 $\pm$ & 0.06 & ... &   3 & ... & 9(0)
\\ 
K242 & ... & 16 15 30.9 & -60 55 49 & 325.31 & -7.27 & 0.072 & C & 16.02 $\pm$ & 0.08 & ... &   2 & ... & 1(0)
\\ 
K243 & ... & 16 15 31.7 & -60 56 54 & 325.30 & -7.28 & 0.072 & G & 14.68 $\pm$ & 0.04 & ... &   4 & ... & 1(0) 8(0)
\\ 
K244 & WKK6305 & 16 15 32.9 & -60 39 55 & 325.50 & -7.08 & 0.098 & G & 9.37 $\pm$ & 0.00 & 14.4 &  38 & 4710 & 10(19) 25(2)
\\ 
K245 & ... & 16 15 33.7 & -60 50 00 & 325.38 & -7.20 & 0.078 & C & 14.76 $\pm$ & 0.03 & ... &   4 & ... & 2(0)
\\ 
K246 & ... & 16 15 35.5 & -60 48 52 & 325.40 & -7.19 & 0.078 & G & 15.06 $\pm$ & 0.04 & ... &   3 & ... & 2(16) 9(0)
\\ 
K247 & R & 16 15 35.5 & -61 12 17 & 325.13 & -7.47 & 0.074 & G & 13.34 $\pm$ & 0.02 & ... &   8 & ... & 37(19) 38(3)
\\ 
K248 & ... & 16 15 35.8 & -60 43 41 & 325.46 & -7.13 & 0.098 & C & 15.87 $\pm$ & 0.06 & ... &   3 & ... & 25(0)
\\ 
K249 & R & 16 15 36.7 & -60 56 45 & 325.31 & -7.29 & 0.072 & G & 13.96 $\pm$ & 0.02 & ... &   6 & 3365 & 8(0)
\\ 
K250 & ... & 16 15 37.7 & -60 59 41 & 325.28 & -7.32 & 0.072 & C & 16.36 $\pm$ & 0.08 & ... &   2 & ... & 7(0)
\\ 
K251 & WKK6309 & 16 15 38.0 & -60 41 59 & 325.49 & -7.11 & 0.098 & G & 12.37 $\pm$ & 0.01 & 16.4 &  10 & 6169 & 25(2)
\\ 
K252 & R & 16 15 40.1 & -60 32 18 & 325.60 & -7.00 & 0.088 & G & 13.74 $\pm$ & 0.01 & ... &   6 & ... & 47(0)
\\ 
K253 & ... & 16 15 40.4 & -61 09 27 & 325.17 & -7.45 & 0.077 & C & 15.40 $\pm$ & 0.04 & ... &   4 & ... & 19(0)
\\ 
K254 & ... & 16 15 42.0 & -61 13 54 & 325.12 & -7.50 & 0.074 & C & 14.97 $\pm$ & 0.03 & ... &   4 & ... & 38(0)
\\ 
K255 & ... & 16 15 42.4 & -61 02 06 & 325.26 & -7.36 & 0.073 & C & 16.58 $\pm$ & 0.08 & ... &   2 & ... & 7(0)
\\ 
K256 & ... & 16 15 42.5 & -61 04 16 & 325.23 & -7.39 & 0.077 & C & 16.41 $\pm$ & 0.06 & ... &   2 & ... & 7(0)
\\ 
K257 & R & 16 15 43.1 & -61 02 38 & 325.25 & -7.37 & 0.073 & G & 16.04 $\pm$ & 0.07 & ... &   2 & ... & 7(0)
\\ 
K258 & ... & 16 15 43.1 & -60 59 13 & 325.29 & -7.33 & 0.072 & C & 16.68 $\pm$ & 0.09 & ... &   2 & ... & 7(0)
\\ 
K259 & WKK6312 & 16 15 46.1 & -60 55 08 & 325.34 & -7.28 & 0.072 & G & 8.54 $\pm$ & 0.00 & 13.1 &  58 & 3839 & 8(2)
\\ 
K260 & R & 16 15 46.3 & -61 08 25 & 325.19 & -7.44 & 0.077 & G & 15.05 $\pm$ & 0.04 & ... &   6 & ... & 19(0)
\\ 
K261 & WKK6313 & 16 15 46.6 & -60 55 22 & 325.34 & -7.29 & 0.072 & G & 11.86 $\pm$ & 0.00 & 16.7 &   8 & 3319 & 8(3)
\\ 
K262 & R & 16 15 47.2 & -60 41 54 & 325.50 & -7.13 & 0.098 & G & 14.04 $\pm$ & 0.03 & ... &   5 & ... & 25(0)
\\ 
K263 & WKK6316 & 16 15 48.0 & -60 42 39 & 325.49 & -7.14 & 0.098 & G & 11.61 $\pm$ & 0.01 & 16.1 &  18 & 4641 & 25(0)
\\ 
K264 & WKK6318 & 16 15 50.2 & -60 48 11 & 325.43 & -7.21 & 0.078 & G & 8.88 $\pm$ & 0.00 & 13.2 &  57 & 3380 & 9(2)
\\ 
K265 & ... & 16 15 51.4 & -60 45 40 & 325.46 & -7.18 & 0.094 & C & 13.47 $\pm$ & 0.01 & ... &   7 & ... & 9(0)
\\ 
K266 & ... & 16 15 52.7 & -61 15 31 & 325.11 & -7.53 & 0.074 & C & 15.22 $\pm$ & 0.04 & ... &   4 & ... & 38(0)
\\ 
K267 & ... & 16 15 52.9 & -60 48 09 & 325.43 & -7.21 & 0.078 & C & 13.80 $\pm$ & 0.02 & ... &   6 & ... & 9(3)
\\ 
K268 & WKK6319 & 16 15 53.0 & -60 50 57 & 325.40 & -7.24 & 0.078 & G & 10.42 $\pm$ & 0.00 & 14.9 &  44 & 3877 & 8(18) 9(2)
\\ 
K269 & ... & 16 15 55.0 & -61 11 33 & 325.16 & -7.49 & 0.073 & G & 14.70 $\pm$ & 0.03 & ... &   5 & ... & 19(0) 38(0)
\\ 
K270 & ... & 16 15 55.2 & -61 10 25 & 325.18 & -7.48 & 0.073 & C & 16.61 $\pm$ & 0.07 & ... &   2 & ... & 19(0)
\\ 
K271 & R & 16 15 55.9 & -61 04 26 & 325.25 & -7.41 & 0.077 & G & 15.43 $\pm$ & 0.03 & ... &   5 & ... & 19(0)
\\ 
K272 & R & 16 15 56.0 & -60 37 19 & 325.57 & -7.08 & 0.095 & G & 15.84 $\pm$ & 0.06 & ... &   3 & ... & 25(0) 47(0)
\\ 
K273 & R & 16 15 57.8 & -60 41 54 & 325.52 & -7.14 & 0.098 & G & 14.79 $\pm$ & 0.04 & ... &   3 & ... & 25(0)
\\ 
K274 & ... & 16 15 58.9 & -61 04 41 & 325.25 & -7.41 & 0.077 & C & 16.73 $\pm$ & 0.08 & ... &   2 & ... & 7(0)
\\ 
K275 & WKK6325 & 16 16 02.6 & -60 57 51 & 325.34 & -7.34 & 0.072 & G & 13.39 $\pm$ & 0.01 & 17.5 &  10 & ... & 7(0) 8(0)
\\ 
K276 & WKK6326 & 16 16 02.7 & -60 39 13 & 325.55 & -7.12 & 0.095 & G & 13.89 $\pm$ & 0.03 & 17.9 &   5 & 5096 & 25(0)
\\ 
K277 & WKK6329 & 16 16 04.6 & -60 30 53 & 325.65 & -7.02 & 0.083 & G & 12.78 $\pm$ & 0.01 & 17.3 &  10 & 4749 & 47(3)
\\ 
K278 & R & 16 16 04.9 & -60 47 47 & 325.46 & -7.22 & 0.094 & G & 15.61 $\pm$ & 0.06 & ... &   4 & ... & 9(0)
\\ 
K279 & ... & 16 16 05.2 & -60 48 47 & 325.45 & -7.23 & 0.078 & C & 16.20 $\pm$ & 0.08 & ... &   2 & ... & 9(0)
\\ 
K280 & WKK6328 & 16 16 06.0 & -60 53 10 & 325.39 & -7.29 & 0.074 & G & 13.08 $\pm$ & 0.01 & 17.5 &  11 & 5870 & 8(0)
\\ 
K281 & R & 16 16 06.4 & -61 10 30 & 325.19 & -7.49 & 0.073 & G & 14.66 $\pm$ & 0.02 & ... &   7 & ... & 19(0)
\\ 
K282 & ... & 16 16 06.4 & -61 14 42 & 325.14 & -7.54 & 0.074 & C & 15.88 $\pm$ & 0.07 & ... &   2 & ... & 38(0)
\\ 
K283 & ... & 16 16 07.5 & -61 04 18 & 325.27 & -7.42 & 0.073 & G & 16.25 $\pm$ & 0.07 & ... &   3 & ... & 7(0)
\\ 
K284 & WKK6360 & 16 16 08.7 & -61 02 36 & 325.29 & -7.40 & 0.073 & G & 11.24 $\pm$ & 0.00 & 16.1 &  19 & 6324 & 7(3)
\\ 
K285 & ... & 16 16 09.3 & -61 12 59 & 325.17 & -7.53 & 0.073 & C & 15.98 $\pm$ & 0.05 & ... &   3 & ... & 38(0)
\\ 
K286 & ... & 16 16 10.3 & -60 39 39 & 325.56 & -7.13 & 0.095 & C & 15.75 $\pm$ & 0.05 & ... &   3 & ... & 25(0)
\\ 
K287 & WKK6336 & 16 16 11.6 & -60 45 07 & 325.50 & -7.20 & 0.094 & G & 13.34 $\pm$ & 0.01 & 17.6 &   7 & ... & 9(0)
\\ 
K288 & R & 16 16 11.9 & -60 38 57 & 325.57 & -7.13 & 0.095 & G & 15.09 $\pm$ & 0.04 & ... &   3 & ... & 25(0)
\\ 
K289 & R & 16 16 13.2 & -60 51 28 & 325.43 & -7.28 & 0.074 & G & 13.95 $\pm$ & 0.02 & ... &   5 & ... & 8(0) 9(0)
\\ 
K290 & R & 16 16 14.0 & -60 51 25 & 325.43 & -7.28 & 0.074 & G & 14.09 $\pm$ & 0.02 & ... &   6 & ... & 8(0) 9(0)
\\ 
K291 & ... & 16 16 15.1 & -60 53 03 & 325.41 & -7.30 & 0.074 & G & 15.98 $\pm$ & 0.06 & ... &   4 & ... & 8(0)
\\ 
K292 & ... & 16 16 15.4 & -61 11 52 & 325.19 & -7.52 & 0.073 & C & 16.09 $\pm$ & 0.05 & ... &   3 & ... & 19(0)
\\ 
K293 & WKK6342 & 16 16 18.9 & -60 57 23 & 325.36 & -7.36 & 0.076 & G & 11.10 $\pm$ & 0.00 & 16.3 &  15 & 4828 & 7(3) 8(3) 21(3) 22(19)
\\ 
K294 & ... & 16 16 19.4 & -60 58 26 & 325.35 & -7.37 & 0.076 & C & 13.95 $\pm$ & 0.01 & ... &   6 & ... & 22(16)
\\ 
K295 & WKK6340 & 16 16 19.5 & -61 17 46 & 325.13 & -7.60 & 0.076 & G & 9.31 $\pm$ & 0.00 & 14.1 &  52 & 6330 & 38(3)
\\ 
K296 & R & 16 16 21.6 & -60 46 58 & 325.49 & -7.24 & 0.094 & G & 13.31 $\pm$ & 0.02 & ... &   7 & ... & 9(0) 23(0)
\\ 
K297 & ... & 16 16 21.9 & -61 03 13 & 325.30 & -7.43 & 0.073 & G & 12.83 $\pm$ & 0.01 & ... &   8 & ... & 7(3) 21(3)
\\ 
K298 & ... & 16 16 22.2 & -60 58 11 & 325.36 & -7.37 & 0.076 & C & 13.86 $\pm$ & 0.02 & ... &   6 & ... & 8(0) 22(0)
\\ 
K299 & R & 16 16 22.3 & -61 17 31 & 325.13 & -7.60 & 0.076 & G & 12.43 $\pm$ & 0.01 & ... &  13 & ... & 38(3) 39(19)
\\ 
K300 & ... & 16 16 22.6 & -60 37 04 & 325.61 & -7.12 & 0.095 & C & 15.75 $\pm$ & 0.07 & ... &   2 & ... & 25(0)
\\
K301 & WKK6349 & 16 16 23.1 & -60 52 03 & 325.43 & -7.30 & 0.074 & G & 13.02 $\pm$ & 0.01 & 16.9 &   8 & 4591 & 8(3) 22(3)
\\ 
K302 & R & 16 16 24.0 & -60 47 03 & 325.49 & -7.24 & 0.094 & G & 12.93 $\pm$ & 0.01 & ... &   7 & ... & 9(0) 23(0)
\\ 
K303 & ... & 16 16 24.7 & -60 40 34 & 325.57 & -7.16 & 0.094 & C & 16.13 $\pm$ & 0.08 & ... &   2 & ... & 25(0)
\\ 
K304 & ... & 16 16 26.0 & -60 35 57 & 325.63 & -7.11 & 0.095 & C & 14.98 $\pm$ & 0.02 & ... &   4 & ... & 46(0) 47(0)
\\ 
K305 & WKK6351 & 16 16 27.0 & -61 14 23 & 325.18 & -7.57 & 0.073 & G & 14.72 $\pm$ & 0.03 & 17.3 &   2 & 4451 & 38(0) 39(2)
\\ 
K306 & ... & 16 16 27.2 & -61 17 11 & 325.14 & -7.60 & 0.076 & G & 14.89 $\pm$ & 0.04 & ... &   4 & ... & 38(0) 39(0)
\\ 
K307 & ... & 16 16 28.9 & -60 50 18 & 325.46 & -7.29 & 0.086 & C & 16.07 $\pm$ & 0.07 & ... &   2 & ... & 9(0)
\\ 
K308 & R & 16 16 31.9 & -61 03 06 & 325.32 & -7.44 & 0.073 & G & 15.13 $\pm$ & 0.03 & ... &   4 & ... & 21(0)
\\ 
K309 & ... & 16 16 32.7 & -61 10 15 & 325.23 & -7.53 & 0.073 & G & 14.52 $\pm$ & 0.03 & ... &   5 & ... & 19(16) 20(0)
\\ 
K310 & ... & 16 16 33.4 & -61 04 27 & 325.30 & -7.46 & 0.076 & C & 16.22 $\pm$ & 0.06 & ... &   2 & ... & 21(0)
\\ 
K311 & ... & 16 16 35.3 & -61 13 55 & 325.19 & -7.58 & 0.073 & C & 15.97 $\pm$ & 0.06 & ... &   3 & ... & 39(0)
\\ 
K312 & R & 16 16 36.2 & -60 50 20 & 325.47 & -7.30 & 0.086 & G & 15.61 $\pm$ & 0.06 & ... &   2 & ... & 23(0)
\\ 
K313 & WKK6359 & 16 16 36.7 & -61 05 25 & 325.30 & -7.48 & 0.076 & G & 13.68 $\pm$ & 0.02 & 17.9 &   8 & 6257 & 20(2)
\\ 
K314 & WKK6360 & 16 16 37.0 & -61 02 45 & 325.33 & -7.44 & 0.073 & G & 10.14 $\pm$ & 0.00 & 15.3 &  27 & 6258 & 21(0)
\\ 
K315 & ... & 16 16 38.5 & -61 07 60 & 325.27 & -7.51 & 0.076 & G & 13.11 $\pm$ & 0.01 & ... &  16 & ... & 20(0)
\\ 
K316 & ... & 16 16 40.2 & -61 09 40 & 325.25 & -7.53 & 0.076 & C & 16.27 $\pm$ & 0.05 & ... &   3 & ... & 20(0)
\\ 
K317 & WKK6364 & 16 16 40.6 & -60 59 53 & 325.37 & -7.42 & 0.076 & G & 12.40 $\pm$ & 0.01 & 17.1 &  11 & ... & 21(0)
\\ 
K318 & ... & 16 16 41.1 & -60 52 13 & 325.46 & -7.33 & 0.086 & C & 15.29 $\pm$ & 0.04 & ... &   3 & ... & 22(0)
\\ 
K319 & R & 16 16 42.6 & -61 01 47 & 325.35 & -7.44 & 0.073 & G & 14.38 $\pm$ & 0.03 & ... &   6 & ... & 21(0)
\\ 
K320 & ... & 16 16 42.8 & -60 52 55 & 325.45 & -7.34 & 0.086 & C & 14.92 $\pm$ & 0.04 & ... &   4 & ... & 22(0)
\\ 
K321 & ... & 16 16 43.4 & -61 06 12 & 325.30 & -7.50 & 0.076 & C & 15.69 $\pm$ & 0.04 & ... &   3 & ... & 20(0)
\\ 
K322 & R & 16 16 46.8 & -61 18 38 & 325.15 & -7.65 & 0.073 & G & 14.75 $\pm$ & 0.04 & ... &   4 & 24499 & 39(0)
\\ 
K323 & ... & 16 16 47.0 & -60 51 14 & 325.48 & -7.32 & 0.086 & G & 15.13 $\pm$ & 0.03 & ... &   4 & ... & 22(0)
\\ 
K324 & ... & 16 16 47.3 & -60 54 02 & 325.44 & -7.36 & 0.076 & C & 16.05 $\pm$ & 0.07 & ... &   3 & ... & 22(0)
\\ 
K325 & WKK6370 & 16 16 49.6 & -61 08 49 & 325.27 & -7.53 & 0.076 & G & 9.91 $\pm$ & 0.00 & 16.9 &  36 & 4702 & 20(2)
\\ 
K326 & ... & 16 16 51.6 & -61 05 45 & 325.31 & -7.50 & 0.076 & C & 15.96 $\pm$ & 0.05 & ... &   3 & ... & 20(0)
\\ 
K327 & ... & 16 16 51.9 & -60 55 09 & 325.44 & -7.38 & 0.076 & C & 15.52 $\pm$ & 0.04 & ... &   3 & ... & 22(0)
\\ 
K328 & R & 16 16 52.0 & -60 43 44 & 325.57 & -7.24 & 0.094 & G & 14.44 $\pm$ & 0.02 & ... &   4 & ... & 24(0)
\\ 
K329 & WKK6375 & 16 16 52.1 & -60 49 25 & 325.51 & -7.31 & 0.086 & G & 15.40 $\pm$ & 0.04 & 17.4 &   2 & 5249 & 23(0)
\\ 
K330 & ... & 16 16 53.1 & -61 12 41 & 325.23 & -7.59 & 0.073 & C & 16.27 $\pm$ & 0.08 & ... &   2 & ... & 39(0)
\\ 
K331 & R & 16 16 55.5 & -61 02 58 & 325.35 & -7.47 & 0.082 & G & 13.98 $\pm$ & 0.02 & ... &   6 & ... & 21(0)
\\ 
K332 & ... & 16 16 55.5 & -61 02 26 & 325.36 & -7.47 & 0.082 & C & 15.88 $\pm$ & 0.04 & ... &   3 & ... & 21(0)
\\ 
K333 & R & 16 16 55.7 & -60 57 41 & 325.41 & -7.41 & 0.076 & G & 14.46 $\pm$ & 0.04 & ... &   5 & ... & 22(3)
\\ 
K334 & ... & 16 16 56.0 & -60 43 44 & 325.58 & -7.25 & 0.094 & G & 16.01 $\pm$ & 0.06 & ... &   3 & ... & 24(0)
\\ 
K335 & ... & 16 16 57.0 & -61 08 08 & 325.29 & -7.54 & 0.076 & G & 14.49 $\pm$ & 0.01 & ... &   5 & ... & 20(0)
\\ 
K336 & ... & 16 16 57.2 & -60 56 33 & 325.43 & -7.40 & 0.076 & C & 15.79 $\pm$ & 0.07 & ... &   2 & ... & 22(0)
\\ 
K337 & WKK6380 & 16 16 57.7 & -60 42 51 & 325.59 & -7.24 & 0.094 & G & 13.38 $\pm$ & 0.01 & 17.7 &  10 & 18800 & 24(0)
\\ 
K338 & ... & 16 16 58.3 & -60 54 03 & 325.46 & -7.37 & 0.084 & C & 15.07 $\pm$ & 0.05 & ... &   4 & ... & 22(0)
\\ 
K339 & R & 16 16 58.3 & -60 41 29 & 325.61 & -7.22 & 0.094 & G & 15.55 $\pm$ & 0.05 & ... &   2 & ... & 24(0)
\\ 
K340 & ... & 16 16 59.1 & -61 09 60 & 325.27 & -7.56 & 0.076 & C & 14.95 $\pm$ & 0.03 & ... &   4 & ... & 20(0)
\\ 
K341 & R & 16 16 59.9 & -61 10 08 & 325.27 & -7.57 & 0.076 & G & 13.58 $\pm$ & 0.01 & ... &   7 & 29841 & 20(0)
\\ 
K342 & ... & 16 17 00.1 & -61 15 17 & 325.21 & -7.63 & 0.073 & C & 14.20 $\pm$ & 0.01 & ... &   6 & ... & 39(0)
\\ 
K343 & WKK6383 & 16 17 00.4 & -60 52 25 & 325.48 & -7.36 & 0.084 & G & 10.80 $\pm$ & 0.00 & 15.9 &  25 & 5431 & 22(2)
\\ 
K344 & ... & 16 17 01.2 & -60 38 03 & 325.65 & -7.19 & 0.092 & C & 15.92 $\pm$ & 0.06 & ... &   3 & ... & 24(0)
\\ 
K345 & WKK6385 & 16 17 01.2 & -60 41 36 & 325.61 & -7.23 & 0.094 & G & 14.53 $\pm$ & 0.03 & 17.2 &   4 & 7141 & 24(0)
\\ 
K346 & R & 16 17 01.4 & -61 09 18 & 325.29 & -7.56 & 0.076 & G & 13.40 $\pm$ & 0.01 & ... &   7 & ... & 20(0)
\\ 
K347 & R & 16 17 02.1 & -60 44 52 & 325.57 & -7.27 & 0.093 & G & 15.05 $\pm$ & 0.03 & ... &   4 & ... & 23(0) 24(16)
\\ 
K348 & R & 16 17 02.6 & -61 07 47 & 325.31 & -7.54 & 0.076 & G & 14.49 $\pm$ & 0.02 & ... &   5 & ... & 20(0)
\\ 
K349 & ... & 16 17 05.2 & -60 52 08 & 325.49 & -7.36 & 0.084 & C & 15.45 $\pm$ & 0.03 & ... &   3 & ... & 22(0)
\\ 
K350 & R & 16 17 06.1 & -60 55 37 & 325.45 & -7.40 & 0.084 & G & 15.69 $\pm$ & 0.06 & ... &   3 & ... & 22(0)
\\ 
K351 & R & 16 17 07.4 & -60 52 29 & 325.49 & -7.37 & 0.084 & G & 14.54 $\pm$ & 0.02 & ... &   5 & ... & 22(0)
\\ 
K352 & ... & 16 17 09.3 & -61 07 01 & 325.32 & -7.54 & 0.076 & C & 16.31 $\pm$ & 0.07 & ... &   3 & ... & 20(0)
\\ 
K353 & ... & 16 17 11.7 & -61 06 21 & 325.34 & -7.54 & 0.076 & G & 15.55 $\pm$ & 0.05 & ... &   4 & ... & 20(0)
\\ 
K354 & ... & 16 17 17.7 & -60 43 58 & 325.61 & -7.28 & 0.093 & C & 14.62 $\pm$ & 0.03 & ... &   6 & ... & 24(0)
\\ 
K355 & ... & 16 17 18.6 & -61 05 13 & 325.36 & -7.53 & 0.082 & C & 15.18 $\pm$ & 0.03 & ... &   5 & ... & 20(0)
\\ 
K356 & ... & 16 17 18.9 & -60 37 54 & 325.68 & -7.21 & 0.092 & G & 14.84 $\pm$ & 0.02 & ... &   4 & ... & 24(0) 45(0) 46(0)
\\ 
K357 & R & 16 17 20.0 & -60 53 11 & 325.50 & -7.39 & 0.084 & G & 13.57 $\pm$ & 0.01 & ... &   7 & ... & 43(3)
\\ 
K358 & R & 16 17 20.4 & -60 51 53 & 325.52 & -7.38 & 0.084 & G & 13.39 $\pm$ & 0.02 & ... &   9 & 3960 & 22(2) 43(2)
\\ 
K359 & WKK6402 & 16 17 22.1 & -60 38 03 & 325.68 & -7.22 & 0.092 & G & 10.58 $\pm$ & 0.00 & 15.4 &  37 & 5826 & 24(2) 45(2) 46(18)
\\ 
K360 & WKK6406 & 16 17 25.0 & -60 37 55 & 325.69 & -7.22 & 0.092 & G & 14.45 $\pm$ & 0.03 & 17.3 &   7 & 7349 & 24(0) 45(0) 46(16)
\\ 
K361 & R & 16 17 25.6 & -61 03 01 & 325.39 & -7.52 & 0.082 & G & 16.31 $\pm$ & 0.08 & ... &   2 & ... & 41(0)
\\ 
K362 & R & 16 17 26.5 & -61 16 29 & 325.24 & -7.68 & 0.078 & G & 14.37 $\pm$ & 0.02 & ... &   5 & 18551 & 39(0)
\\ 
K363 & WKK6407 & 16 17 27.4 & -61 02 59 & 325.40 & -7.52 & 0.082 & G & 14.07 $\pm$ & 0.03 & 18.1 &   7 & 3653 & 21(16) 41(0)
\\ 
K364 & R & 16 17 29.6 & -60 53 34 & 325.51 & -7.41 & 0.084 & G & 13.78 $\pm$ & 0.01 & ... &   5 & ... & 43(0)
\\ 
K365 & ... & 16 17 32.7 & -60 53 43 & 325.52 & -7.42 & 0.084 & G & 12.99 $\pm$ & 0.01 & ... &   8 & ... & 42(2) 43(0)
\\ 
K366 & R & 16 17 34.6 & -60 42 02 & 325.66 & -7.28 & 0.097 & G & 15.53 $\pm$ & 0.05 & ... &   4 & ... & 44(0)
\\ 
K367 & ... & 16 17 37.1 & -60 54 38 & 325.51 & -7.43 & 0.084 & C & 14.35 $\pm$ & 0.02 & ... &   5 & ... & 43(0)
\\ 
K368 & WKK6416 & 16 17 38.3 & -61 08 36 & 325.35 & -7.60 & 0.076 & G & 14.57 $\pm$ & 0.04 & 18.0 &   5 & ... & 40(0) 41(16)
\\ 
K369 & ... & 16 17 40.1 & -60 41 56 & 325.66 & -7.29 & 0.097 & C & 15.99 $\pm$ & 0.05 & ... &   3 & ... & 44(0)
\\ 
K370 & WKK6419 & 16 17 43.7 & -60 38 52 & 325.71 & -7.26 & 0.097 & G & 14.08 $\pm$ & 0.03 & 17.7 &   6 & 4453 & 45(0)
\\ 
K371 & ... & 16 17 45.5 & -60 39 36 & 325.70 & -7.27 & 0.097 & G & 15.88 $\pm$ & 0.06 & ... &   3 & ... & 45(0)
\\ 
K372 & ... & 16 17 45.8 & -60 56 06 & 325.51 & -7.46 & 0.088 & G & 15.33 $\pm$ & 0.04 & ... &   3 & ... & 42(0)
\\ 
K373 & ... & 16 17 48.4 & -61 09 49 & 325.35 & -7.63 & 0.076 & C & 15.29 $\pm$ & 0.05 & ... &   3 & ... & 40(0)
\\ 
K374 & R & 16 17 50.0 & -61 10 04 & 325.35 & -7.63 & 0.076 & G & 15.43 $\pm$ & 0.05 & ... &   2 & ... & 40(0)
\\ 
K375 & ... & 16 17 53.6 & -61 02 33 & 325.44 & -7.55 & 0.085 & C & 15.51 $\pm$ & 0.03 & ... &   4 & ... & 41(0)
\\ 
K376 & WKK6429 & 16 17 54.9 & -61 12 59 & 325.32 & -7.68 & 0.076 & G & 11.50 $\pm$ & 0.01 & 16.4 &  16 & 4035 & 40(3)
\\ 
K377 & R & 16 17 56.2 & -60 55 38 & 325.53 & -7.47 & 0.088 & G & 13.91 $\pm$ & 0.02 & ... &   7 & ... & 42(0)
\\ 
K378 & R & 16 17 56.2 & -61 05 57 & 325.40 & -7.59 & 0.085 & G & 14.09 $\pm$ & 0.02 & ... &   8 & ... & 41(0)
\\ 
K379 & WKK6430 & 16 17 56.8 & -61 08 00 & 325.38 & -7.62 & 0.085 & G & 13.38 $\pm$ & 0.01 & 17.3 &   8 & ... & 40(0) 41(0)
\\ 
K380 & WKK6431 & 16 17 57.3 & -60 55 23 & 325.53 & -7.47 & 0.088 & G & 10.90 $\pm$ & 0.00 & 15.7 &  20 & 3333 & 42(2)
\\ 
K381 & R & 16 17 59.0 & -60 55 05 & 325.54 & -7.47 & 0.088 & G & 14.46 $\pm$ & 0.03 & ... &   4 & ... & 42(0)
\\ 
K382 & R & 16 17 59.0 & -61 07 15 & 325.39 & -7.61 & 0.085 & G & 15.02 $\pm$ & 0.04 & ... &   3 & ... & 41(0)
\\ 
K383 & R & 16 18 00.6 & -60 41 19 & 325.70 & -7.31 & 0.097 & G & 15.06 $\pm$ & 0.04 & ... &   4 & ... & 44(0)
\\ 
K384 & ... & 16 18 00.8 & -60 37 30 & 325.75 & -7.26 & 0.097 & G & 15.48 $\pm$ & 0.05 & ... &   3 & ... & 45(0)
\\ 
K385 & ... & 16 18 01.3 & -60 40 36 & 325.71 & -7.30 & 0.097 & C & 15.33 $\pm$ & 0.03 & ... &   3 & ... & 45(0)
\\ 
K386 & WKK6439 & 16 18 04.2 & -60 41 42 & 325.70 & -7.32 & 0.097 & G & 10.27 $\pm$ & 0.00 & 15.3 &  31 & 3942 & 44(2)
\\ 
K387 & WKK6443 & 16 18 07.4 & -60 56 48 & 325.53 & -7.50 & 0.088 & G & 15.49 $\pm$ & 0.05 & 18.1 &   2 & ... & 42(0)
\\ 
K388 & ... & 16 18 08.1 & -60 47 48 & 325.64 & -7.40 & 0.093 & C & 16.09 $\pm$ & 0.07 & ... &   2 & ... & 44(0)
\\ 
K389 & ... & 16 18 10.0 & -61 02 28 & 325.47 & -7.57 & 0.089 & C & 15.71 $\pm$ & 0.06 & ... &   3 & ... & 41(0)
\\ 
K390 & ... & 16 18 11.1 & -61 07 60 & 325.40 & -7.64 & 0.085 & C & 16.07 $\pm$ & 0.07 & ... &   3 & ... & 41(0)
\\ 

\end{longtable}
\label{lastpage}

\end{document}